\documentclass[aos,MSNbibl,seceqn,nameyear,dvips]{arximspdf}
\usepackage{graphicx}

\doi{10.1214/11-AOS939}
\volume{39}
\issue{6}
\pubyear{2011}
\firstpage{3121}
\lastpage{3151}

\makeatletter

\newtheorem{theorem}{Theorem}
\newtheorem{cor}{Corollary}

\newproclaim{dfn}{Definition}

\newtheorem{lemma}{Lemma}
\newtheorem{prop}{Proposition}

\newtheorem{claim}{Claim}
\newproclaim{remark}{Remark}

\newcommand{\tr}{\operatorname{tr}}
\newcommand{\re}{\operatorname{Re}}
\newcommand{\im}{\operatorname{Im}}

\newcommand{\zz}[1]{\mathbb{#1}}
\newcommand{\la}{\langle}
\newcommand{\ra}{\rangle}

\newcommand{\sC}{\mathcal{C}}
\newcommand{\sI}{\mathcal{I}}

\newcommand{\wt}{\widetilde}
\newcommand{\wh}{\widehat}

\newcommand{\vep}{\varepsilon}
\newcommand{\ol}{\overline}
\newcommand{\eqd}{\stackrel{d}{=}}

\makeatother

\begin{document}
\begin{frontmatter}

\title{On the estimation of integrated covariance matrices of high
dimensional diffusion processes\thanksref{T1}}
\runtitle{High dimensional integrated covariance matrices}

\thankstext{T1}{Supported in part by DAG (HKUST) and GRF 606811 of the HKSAR.}

\begin{aug}
\author[A]{\fnms{Xinghua} \snm{Zheng}\corref{}\ead[label=e1]{xhzheng@ust.hk}}
\and
\author[A]{\fnms{Yingying} \snm{Li}\ead[label=e2]{yyli@ust.hk}}
\runauthor{X. Zheng and Y. Li}
\affiliation{Hong Kong University of Science and Technology}
\address[A]{Department of Information Systems\\
\quad Business Statistics and Operations Management\\
Hong Kong University of Science \\
\quad and Technology\\
Clear Water Bay, Kowloon\\
Hong Kong\\
\printead{e1}\\
\hphantom{E-mail: }\printead*{e2}} 
\end{aug}

\received{\smonth{5} \syear{2010}}
\revised{\smonth{10} \syear{2011}}

%
\begin{abstract}
We consider the estimation of integrated covariance (ICV) matrices of
high dimensional diffusion processes based on high frequency
observations. We start by studying the most commonly used estimator,
the \textit{realized covariance} (RCV) matrix. We show that in the high
dimensional case when the dimension $p$ and the observation
frequency~$n$ grow in the same rate, the limiting spectral distribution
(LSD) of RCV depends on the covolatility process \textit{not only through
the targeting ICV}, \textit{but also on how the covolatility process
varies in time}. We establish a Mar\v{c}enko--Pastur type theorem for
weighted sample covariance matrices, based on which we obtain a~Mar\v{c}enko--Pastur type theorem for RCV for a class $\sC$ of diffusion
processes. The results explicitly demonstrate how the time variability
of the covolatility process affects the LSD of RCV. We further propose
an alternative estimator, the \textit{time-variation adjusted realized
covariance} (TVARCV) matrix. We show that for processes in class $\sC$,
the TVARCV possesses the desirable property that its LSD depends
solely on that of the targeting ICV through the Mar\v{c}enko--Pastur
equation, and hence, in particular, the TVARCV can be used to recover
the empirical spectral distribution of the ICV by using existing
algorithms.
\end{abstract}

%
\begin{keyword}[class=AMS]
\kwd[Primary ]{62H12}
\kwd[; secondary ]{62G99}
\kwd[; tertiary ]{60F15}.
\end{keyword}
\begin{keyword}
\kwd{High dimension}
\kwd{high frequency}
\kwd{integrated covariance matrix}
\kwd{Mar\v{c}enko--Pastur equation}
\kwd{weighted sample covariance matrix}
\kwd{realized covariance matrix}.
\end{keyword}

\end{frontmatter}

\section{Introduction}\label{secintro}

\subsection{Background}

Diffusion processes are widely used to model financial asset price
processes.
For example, suppose that we have multiple stocks, say, $p$ stocks whose
price processes are denoted by $S^{(j)}_t$ for $j=1,\ldots,p$, and
$X_t^{(j)}:=\log S_t^{(j)}$ are the log price processes. Let
$\mathbf{X}_t = (X_t^{(1)},\ldots,X_t^{(p)})^T$. Then a widely used
model for $\mathbf{X}_t$ is [see, e.g., Definition 1 in \citet{BNS04}]
%
\begin{equation}\label{eqX}
d \mathbf{X}_t = {\bolds\mu}_t  \,dt + \Theta_t  \,d\mathbf{W}_t,
\end{equation}
where, $\bolds{\mu}_t=(\mu_t^{(1)},\ldots,\mu_t^{(p)})^T$ is a $p$-dimensional
drift process; $\Theta_t$ is a \mbox{$p\times p$} matrix for any $t$, and
is called the (instantaneous) \textit{covolatility process}; and~$\mathbf{W}_t$ is a $p$-dimensional standard Brownian motion.

The \textit{integrated covariance} (ICV) matrix
\[
{ \Sigma_p
:=\int_0^1 \Theta_t \Theta_t^T  \,dt }
\]
is of great interest in financial applications, which
in the one dimensional case is known as the \textit{integrated volatility}.
A widely used estimator of the ICV matrix is the so-called \textit{realized
covariance} (RCV) matrix, which is defined as follows. Assume that
we can observe the processes $X^{(j)}_t$'s at high frequency synchronously,
say, at time points $\tau_{n,\ell}$:
\[
X^{(j)}_{\tau_{n,\ell}} \bigl(\mbox{$=$}\log S^{(j)}_{\tau_{n,\ell}}\bigr),\qquad
\ell =0,1,\ldots,n,
j=1,\ldots,p,
\]
then the RCV matrix is defined as
%
\begin{eqnarray}\label{eqRCV}
\Sigma^{\mathrm{RCV}}_p
:=\sum_{\ell=1}^n \Delta\mathbf{X}_{\ell}
(\Delta\mathbf{X}_{\ell})^T\qquad\nonumber\\[-8pt]\\[-8pt]
&&\eqntext{\mbox{where } \Delta\mathbf{X}_{\ell}
=\pmatrix{
\Delta X^{(1)}_\ell\cr
\vdots\cr
\Delta X^{(p)}_\ell}
:=\pmatrix{
X^{(1)}_{\tau_{n,\ell}}-X^{(1)}_{\tau_{n,\ell-1}}\cr
\vdots\cr
X^{(p)}_{\tau_{n,\ell}}-X^{(p)}_{\tau_{n,\ell-1}}}.}
\end{eqnarray}
In the one dimensional case, the RCV matrix reduces to the \textit
{realized volatility}.
Thanks to its nice convergence to the ICV matrix as the observation
frequency $n$ goes to infinity [see \citet{jp98}], the RCV matrix is
highly appreciated in both academic research and practical
applications.
%
\begin{remark}
The tick-by-tick data are usually not observed synchronous\-ly, and
moreover are contaminated by market microstructure noise. On sparsely
sampled data (e.g., 5-minute data for some highly liquid assets, or
subsample from data synchronized by refresh times [\citet{BHLS11}]),
the theory in this paper should be readily applicable, just as one can
use the realized volatility based on sparsely sampled data to estimate
the integrated volatility; see, for example, \citet{ABDL01}.
\end{remark}

\subsection{Large dimensional random
matrix theory (LDRMT)}\label{ssecldrmt}

Having a good estimate of the ICV matrix $\Sigma_p$, in particular,
its spectrum (i.e., its set of eigenvalues\vadjust{\goodbreak}
$\{\lambda_j\dvtx j=1,\ldots,p\}$), is crucial in many applications
such as principal component analysis and portfolio optimization (see,
e.g., the pioneer work of
Markowitz (\citeyear{Markowitz52,Markowitz59}) and a more recent work
[\citet{BaiLW09}]). When the dimension $p$ is high, it is more
convenient to study, instead of the $p$ eigenvalues
$\{\lambda_j\dvtx j=1,\ldots,p\}$, the associated \textit{empirical
spectral distribution} (ESD)
\[
F^{\Sigma_p}(x):=\frac{1}{p}\#\{j\dvtx \lambda_j\leq
x\}, \qquad x\in\zz{R}.
\]
A naive estimator of the spectrum of the ICV matrix $\Sigma_p$ is
the spectrum of the RCV matrix
$\Sigma^{\mathrm{RCV}}_p$. In particular, one wishes that the ESD $F^{\Sigma
^{\mathrm{RCV}}_p}$ of~$\Sigma^{\mathrm{RCV}}_p$
would approximate $F^{\Sigma_p}$ well when the frequency $n$ is
sufficiently high. From the large dimensional random
matrix theory (LDRMT), we now understand quite well that in the high
dimensional setting this good
wish won't come true. For example, in the simplest case when the
drift process is 0, covolatility process is
constant, and observation times $\tau_{n,\ell}$ are equally
spaced, namely, $\tau_{n,\ell} = \ell/n$, we are in the setting of
estimating the usual \textit{covariance matrix} using the
\textit{sample covariance matrix}, given $n$ i.i.d.
$p$-dimensional observations $(\Delta\mathbf{X}_\ell)_{\ell=1,\ldots
,n}$. From LDRMT, we know that if~$p/n$
converges to a~non-zero number and the ESD $F^{\Sigma_p}$ of the
true covariance matrix converges, then the ESD
$F^{\Sigma^{\mathrm{RCV}}_p}$ of the sample covariance matrix also
converges; see, for example, \citet{MP67}, \citet{Yin86}, \citet{SB95}
and \citet{Silverstein95}. The relationship
between the \textit{limiting spectral distribution} (LSD) of $\Sigma
^{\mathrm{RCV}}_p$ in this case\vspace*{1pt} and the
LSD of $\Sigma_p$ can be described by a Mar\v{c}enko--Pastur
equation through Stieltjes transforms, as follows.
%
\begin{prop}[{[Theorem 1.1 of \citet{Silverstein95}]}]\label{propMPgen}
Assume on a common probability space:
\begin{longlist}
\item for $p=1,2,\ldots$ and for $1\leq\ell\leq n$, $\mathbf{Z}^{(p)}_\ell=(Z^{(p,j)}_{\ell})_{1\leq j\leq p}$ with
$Z^{(p,j)}_{\ell}$ i.i.d. with mean 0 and variance 1;
\item\label{asmyn}
$n=n(p)$ with $y_n:=p/n\rightarrow y>0$ as $p\rightarrow\infty$;
\item\label{asmconvSigma}
$\Sigma_p$ is a (possibly random)
nonnegative definite $p\times p$ matrix such that its ESD $F^{\Sigma
_p}$ converges almost surely in distribution to a probability
distribution $H$ on $[0,\infty)$ as $p\rightarrow\infty$;
\item$\Sigma_p$ and $\mathbf{Z}^{(p)}_\ell$'s are independent.
\end{longlist}
Let $\Sigma_p^{1/2}$ be the (nonnegative) square root matrix of
$\Sigma_p$ and
$S_p:= 1/n\times\sum_{\ell=1}^n \Sigma_p^{1/2}\mathbf{Z}^{(p)}_\ell(\mathbf{Z}^{(p)}_\ell)^T \Sigma_p^{1/2}$. Then, almost surely, the ESD of
$S_p$ converges in distribution to a probability distribution $F$,
which is
determined by $H$ in
that its Stieltjes transform
\[
m_F(z):=\int_{\lambda\in\zz{R}} \frac{1}{\lambda- z}  \,dF(\lambda
), \qquad  z\in
\zz{C}_{+}:=\{z\in\zz{C}\dvtx \im(z)>0\}
\]
solves the equation
%
\begin{equation}\label{eqSTFH0}
m_F(z)= \int_{\tau\in\zz{R}} \frac{1}{\tau(1-y(1+zm_F(z))) -z}\,
dH(\tau).
\end{equation}
\end{prop}

In the special case when $\Sigma_p= \sigma^2 I_{p\times p}$, where
$I_{p\times p}$ is the $p\times p$ identity matrix, the LSD $F$ can be
explicitly expressed as follows.
%
\begin{prop}[{[see, e.g., Theorem 2.5 in \citet{Bai99}]}]\label
{propMP} Suppose that~$\mathbf{Z}^{(p)}_\ell$'s are as in
the previous proposition, and $\Sigma_p= \sigma^2 I_{p\times p}$ for
some $\sigma^2>0$. Then
the LSD $F$ has density
\[
p(x) = \frac{1}{2\pi\sigma^2
xy}\sqrt{(b-x)(x-a)}\qquad \mbox{if } a\leq x\leq b,
\]
and a point mass $1-1/y$ at the origin if $y>1$, where
%
\begin{equation}\label{eqMPSupp}
a=a(y)=\sigma^2\bigl(1-\sqrt{y}\bigr)^2\quad \mbox{and}\quad
b=b(y)=\sigma^2\bigl(1+\sqrt{y}\bigr)^2.
\end{equation}
\end{prop}

The LSD $F$ in this proposition is called the
Mar\v{c}enko--Pastur law with ratio index $y$ and scale index
$\sigma^2$, and will be denoted by MP$^{(y,\sigma^2)}$ in this
article.

\subsection{Back to the stochastic volatility case}

In practice, the covolatility process
is typically not constant. For example, it is commonly
observed that
the stock intraday volatility tends to be U-shaped [see, e.g.,
\citet{AP88}, \citet{AB97}]
or exhibits some other patterns [see, e.g., \citet{AB98}].
In this article, we shall allow them to be not only
varying in time but also stochastic. Furthermore, we shall allow the
observation times $\tau_{n,\ell}$ to be random. These
generalizations make our study to be different in nature from the
LDRMT: in LDRMT the observations are i.i.d.; in our
setting, the observations
$(\Delta\mathbf{X}_\ell)_{\ell=1,\ldots,n}$
may, first, be dependant with each other, and second, have
different distributions because (i) the covolatility process may
vary over time, and (ii) the observation durations $\Delta
\tau_\ell:=\tau_{n,\ell} - \tau_{n,\ell-1}$ may be different.

In general, for any time-varying covolatility process $\Theta_t$,
we associate it with a constant covolatility process given by the
square root of the ICV matrix
%
\begin{equation}\label{eqTheta0}
\Theta^0_t:=\sqrt{\int_0^1 \Theta_s \Theta_s^T
ds}\qquad \mbox{for all } t\in[0,1].
\end{equation}
Let $\mathbf{X}_t^0$ be defined by replacing $\Theta_t$ with the constant
covolatility process~$ \Theta^0_t$ (and replacing $\mu_t$ with 0, and
$\mathbf{W}_t$ with another
independent Brownian motion, if necessary) in (\ref{eqX}). Observe
that $\mathbf{X}_t$ and $\mathbf{X}_t^0$ share the same ICV matrix at time 1. Based
on $\mathbf{X}_t^0$, we have an associated RCV matrix
%
\begin{equation}\label{eqRCV0}
\Sigma^{\mathrm{RCV}^0}_p
=\sum_{\ell=1}^n \Delta\mathbf{X}^0_{\ell} (\Delta\mathbf{X}^0_{\ell})^T,
\end{equation}
which is estimating the same ICV matrix as
$\Sigma^{\mathrm{RCV}}_p $.

Since $\Sigma^{\mathrm{RCV}}_p $ and $\Sigma^{\mathrm{RCV}^0}_p $ are based on the
same estimation method and share the same
targeting ICV matrix, it is desirable that their ESDs have similar
properties. In particular, based on the results in LDRMT and the
discussion about constant covolatility case in Section
\ref{ssecldrmt}, we have\vspace*{-2pt} the following property for
$\Sigma^{\mathrm{RCV}^0}_p $: if the ESD $F^{\Sigma_p}$ converges, then so does
$F^{\Sigma^{\mathrm{RCV}^0}_p}$; moreover, their limits are related to
each other via the Mar\v{c}enko--Pastur equation~(\ref{eqSTFH0}).
Does this property also hold for $\Sigma^{\mathrm{RCV}}_p$?
Our first result (Proposition~\ref{propnotconv}) shows that even in
the most ideal case when the covolatility process has the form
$\Theta_t=\gamma_t\cdot I_{p\times p}$ for some deterministic
(scalar) function~$\gamma_t$, such convergence results may
\textit{not} hold for $\Sigma^{\mathrm{RCV}}_p$. In particular,\vspace*{-1pt} the limit of
$F^{\Sigma^{\mathrm{RCV}}_p}$
(when it exists) changes according to how the covolatility process
evolves over time.

This leads to the following natural and interesting question: how does
the LSD of RCV matrix depend on
the time-variability of the covolatility process? Answering this
question in a general context
without putting any structural assumption on the covolatility process
seems to be rather challenging,
if not impossible. For a class $\mathcal{C}$ (see Section \ref
{sectheory}) of processes, we do establish
a result for RCV matrices that's analogous to the Mar\v{c}enko--Pastur
theorem (see Proposition \ref{propMPRCV}),
which demonstrates clearly how the time-variability of the covolatility
process affects the LSD of RCV matrix.
Proposition \ref{propMPRCV} is proved based on Theorem
\ref{thmMPWeightedCov}, which is a Mar\v{c}enko--Pastur type
theorem for \textit{weighted} sample covariance matrices.
These results, in principle, allow one to recover the LSD of ICV
matrix based on that of RCV matrix.

Estimating high dimensional ICV matrices based on high frequency data
has only recently started to gain attention. See, for example,
\citet{WangZou10}; \citet{Taoetal11} who made use of data over long time
horizons by proposing a method incorporating low-frequency dynamics;
and \citet{FanLiYu} who studied the estimation of ICV matrices
for portfolio allocation under gross exposure constraint.
In \citet{WangZou10}, under sparsity assumptions on the ICV matrix,
banding/thresholding was innovatively used to
construct consistent estimators of the ICV matrix in the spectral norm sense.
In particular, when the sparsity assumptions are satisfied, their
estimators share the same LSD as the ICV matrix.
It remains an open question that when the sparsity assumptions are not
satisfied,
whether one can still make good inference about the spectrum of ICV matrix.
For processes in class $\mathcal{C}$ (see
Section \ref{sectheory}), whose ICV matrices do not need to be
sparse, we propose a new estimator, the \textit{time-variation adjusted
realized covariance} (TVARCV) matrix. We show that the TVARCV
matrix has the desirable property that its LSD exists provided
that the LSD of ICV matrix exists, and furthermore, the two LSDs
are related to each other via the Mar\v{c}enko--Pastur equation
(\ref{eqSTFH0}) (see Theorem \ref{thmconvedf}). Therefore, the
TVARCV matrix can be used, for example, to
recover
the LSD of
ICV matrix by inverting the Mar\v{c}enko--Pastur equation
using existing algorithms.

The rest of the paper is organized as the following: theoretical
results are presented in Section \ref{sectheory}, proofs are
given in Section \ref{secproofs}, simulation studies in Section
\ref{secsimulation}, and conclusion and discussions in Section
\ref{secconclusion}.

\textit{Notation.}
For any matrix $A$, $\|A\|=\sqrt{\lambda_{\max}(AA^*)}$ denotes its
spectral norm. For any Hermitian matrix
$A$, $F^A$ stands for its ESD. For two matrices $A$ and $B$, we write
$A\leq B$ ($A\geq B$, resp.) if $B-A$ ($A- B$, resp.) is
a nonnegative definite matrix.
For any interval $I\subseteq[0,\infty)$, and any metric space $S$,
$D(I;S)$ stands for the space of
c\`{a}dl\`{a}g functions from $I$ to $S$. Additionally, $i=\sqrt{-1}$
stands for the
imaginary unit, and for any $z\in\zz{C}$, we write $\re(z), \im(z)$
as its real part
and imaginary part, respectively, and $\ol{z}$ as its complex
conjugate. We also denote $\zz{R}_+=\{a\in\zz{R}\dvtx a>0\}$,
$\zz{C}_+=\{z\in\zz{C}\dvtx\allowbreak \re(z)> 0\}$ and
$Q_1=\{z\in\zz{C}\dvtx \re(z)\geq 0,\im(z) \geq 0\}$. We follow the
custom of writing $f\sim g$ to
mean that the ratio $f/g$ converges to 1.
Finally, throughout the paper, $c,
C,C_1, C'$ etc. denote generic constants whose values may
change from line to line.

\section{Main results}\label{sectheory}

\subsection{Dependance of the LSD of RCV matrix on the
time-variability of covolatility process}

Proposition \ref{propMPgen} asserts that the ESD of sample
covariance matrix
converges to a limiting distribution
which is uniquely determined by the LSD of the underlying covariance matrix.
Unfortunately, Proposition \ref{propMPgen} does not apply to our
case, since
the observations $\Delta\mathbf{X}_\ell$ under our general diffusion
process setting are not i.i.d.
Proposition \ref{propnotconv} below shows that even in the
following most ideal case,
the RCV matrix does not
have the desired convergence property.
%
\begin{prop}\label{propnotconv}
Suppose that for all $p$, $\mathbf{X}_t=\mathbf{X}^{(p)}_t$ is a
$p$-dimensional process satisfying
%
\begin{equation}\label{eqXsimplest}
d \mathbf{X}_t = \gamma_t   \,d\mathbf{W}_t,\qquad t\in[0,1],
\end{equation}
where $\gamma_t> 0$ is a nonrandom (scalar) c\`{a}dl\`{a}g process.
Let $\sigma^2=\int_0^1 \gamma_t^2 \,dt$, and so that the ICV matrix
$\Sigma_p$ is $\sigma^2 I_{p\times p}$. Assume further that
the observation times $\tau_{n,\ell}$ are equally spaced, that is,
$\tau_{n,\ell} = \ell/n$, and that the RCV\vadjust{\goodbreak} matrix~$\Sigma^{\mathrm{RCV}}_p$ is defined by (\ref{eqRCV}). Then so long as
$\gamma_t $ is not constant on $[0,1)$, for any $\vep>0$,
there exists $y_c=y_c(\gamma,\vep)>0$ such that if $\lim p/n = y
\geq y_c$,
%
\begin{equation}\label{eqRCVspt}
\limsup F^{\Sigma^{\mathrm{RCV}}_p}\bigl(b(y)+\sigma^2\vep\bigr) <1\qquad
\mbox{almost surely}.
\end{equation}
In particular, $F^{\Sigma^{\mathrm{RCV}}_p}$ does not converge to the
Mar\v{c}enko--Pastur law MP$^{(y,\sigma^2)}$.
\end{prop}

Observe that MP$^{(y,\sigma^2)}$ is the LSD of RCV matrix when $\gamma
_t\equiv\sigma$. The main message of Proposition \ref{propnotconv} is that, the LSD of RCV matrix depends on the whole
covolatility process \textit{not only through $\Sigma_p$},
\textit{but also on how the covolatility process varies in time}. It
will also be
clear from the proof of Proposition~\ref{propnotconv} (Section
\ref{seccountereg}) that, the more
``volatile'' the covolatility process is, the further away the
LSD is from the Mar\v{c}enko--Pastur law MP$^{(y,\sigma^2)}$.
This is also illustrated in the
simulation study in Section \ref{secsimulation}.

\subsection{The class $\sC$}

To understand the behavior of the ESD of RCV matrix more clearly,
we next focus on a special class of diffusion processes for which we
(i) establish a Mar\v{c}enko--Pastur type theorem for RCV
matrices; and (ii) propose an alternative estimator of ICV matrix.
%
\begin{dfn}
Suppose that $\mathbf{X}_t$
is a $p$-dimensional process satisfying~(\ref{eqX}),
and $\Theta_t $ is c\`{a}dl\`{a}g. We say that $\mathbf{X}_t$ belongs
to class $\mathcal{C}$ if, almost surely, there
exist $(\gamma_t)\in D([0,1];\zz{R})$ and $\Lambda$ a $p\times p$
matrix satisfying $\tr(\Lambda\Lambda^T)=p$ such that
%
\begin{equation}\label{eqclassC}
\Theta_t = \gamma_t \Lambda.
\end{equation}
\end{dfn}

Observe that if (\ref{eqclassC}) holds, then the ICV matrix
$ \Sigma_p = \int_0^1 \gamma_t^2 \,dt\cdot\Lambda\Lambda^T$.
We note that $\Lambda$ does not need to be sparse, hence
neither does $\Sigma_p$.

A special case is when $\Lambda=I_{p\times p}$. This type of
process is studied in Proposition~\ref{propnotconv} and in the
simulation studies in Section \ref{secsimulation}.

A more interesting case is the following.
%
\begin{prop}\label{propsubclassc} Suppose that $X_t^{(j)}$ satisfy
%
\begin{equation}\label{eqX2}
d X_t^{(j)} = \mu_t^{(j)}  \,dt + \sigma_t^{(j)}
dW_t^{(j)}, \qquad j=1,\ldots,p,
\end{equation}
where\vspace*{1pt} $\mu_t^{(j)}, \sigma_t^{(j)}\in D([0,1];\zz{R})$ are the
drift and volatility processes for stock~$j$, and $W_t^{(j)}$'s
are (one-dimensional) standard Brownian motions. If the following
conditions hold:
\begin{longlist}
\item the \textit{correlation matrix} process of $(W_t^{(j)})$
%
\begin{equation}\label{eqcorrBM}
R_t:= \biggl(\frac{\la W^{(j)},W^{(k)}\ra_t}{t}\biggr)_{1\leq
j,k\leq p}
:= \bigl(r^{(jk)}\bigr)_{1\leq j,k\leq p}
\end{equation}
is constant in $t\in(0,1]$;
\item\label{eqrpos} $r^{(jk)} \neq0$ for all $1\leq j,k\leq
p$;  and
\item the correlation matrix process of $(X_t^{(j)})$
%
\begin{equation}\label{eqcorrX}
\biggl(\frac{\int_0^t \sigma_s^{(j)} \sigma_s^{(k)} \,d\la
W^{(j)},W^{(k)}\ra_s}
{\sqrt{\int_0^t (\sigma_s^{(j)})^2 \,ds\cdot\int_0^t (\sigma
_s^{(k)})^2 \,ds}}\biggr)_{1\leq j,k\leq p}
:= \bigl(\rho^{(jk)}\bigr)_{1\leq j,k\leq p}
\end{equation}
is constant in $t\in(0,1]$;
\end{longlist}
then $(X_t^{(j)})$ belongs to class $\mathcal{C}$.
\end{prop}

The proof is given in the supplementary article [\citet{ZL10supp}].

Equation (\ref{eqX2}) is another common way of representing
multi-dimensional log-price processes.
We note that if $X_t^{(j)}$ are
log price processes, then over short time period, say, one day, it
is reasonable to assume that the correlation structure of
$(X_t^{(j)})$ does not change, hence by this proposition,
$(X_t^{(j)})$ belongs to class $\mathcal{C}$.

Observe that if a diffusion process $\mathbf{X}_t$
belongs to class $\mathcal{C}$,
the drift process \mbox{${\bolds\mu}_t\equiv0$},
and $\tau_{n,\ell}$'s and $\gamma_t$ are independent of $\mathbf{W}_t$, then
\[
\Delta\mathbf{X}_{\ell}
=\int_{\tau_{n,\ell-1}}^{\tau_{n,\ell}} \gamma_t \Lambda  \,d\mathbf{W}_t
\eqd\sqrt{\int_{\tau_{n,\ell-1}}^{\tau_{n,\ell}} \gamma_t^2
\,dt}\cdot\breve{\Sigma}^{1/2}\cdot\mathbf{Z}_\ell,
\]
where ``$\eqd$'' stands for ``equal in distribution,'' $\breve{\Sigma
}^{1/2}$ is the nonnegative square root matrix of
$\breve{\Sigma}:=\Lambda\Lambda^T$,
and $\mathbf{Z}_\ell=(Z_\ell^{(1)},\ldots,Z_\ell^{(p)})^T$ consists of
independent
standard normals. Therefore the RCV matrix
\[
\Sigma_p^{\mathrm{RCV}}=\sum_{\ell=1}^n \Delta\mathbf{X}_{\ell} (\Delta\mathbf{X}_{\ell})^T
\eqd\sum_{\ell=1}^n w^n_\ell\cdot\breve{\Sigma}^{1/2} \mathbf{Z}_\ell(\mathbf{Z}_\ell)^T \breve{\Sigma}^{1/2},
\]
where $ w^n_\ell= \int_{\tau_{n,\ell-1}}^{\tau_{n,\ell}} \gamma
_t^2 \,dt$. This is similar to the $S_p$ in Proposition \ref
{propMPgen}, except that here the ``weights'' $w^n_\ell$ may vary in
$\ell$, while in Proposition \ref{propMPgen} the ``weights'' are
constantly $1/n$. Motivated by this observation we develop the
following Mar\v{c}enko--Pastur type theorems for weighted sample
covariance matrices and RCV matrices.

\subsection{Mar\v{c}enko--Pastur type theorems for weighted sample
covariance matrices and RCV matrices}

\begin{theorem}\label{thmMPWeightedCov}
Suppose that assumptions \textup{(ii)} and \textup{(iv)}
in Proposition \ref
{propMPgen} hold. Assume further that:
\begin{longlist}[(A.iii$'$\hspace*{1pt})]
\item[(A.i$'$)] for $p=1,2,\ldots$ and $1\leq\ell\leq n$, $\mathbf{Z}^{(p)}_\ell=(Z^{(p,j)}_{\ell})_{1\leq j\leq p}$ with
$Z^{(p,j)}_{\ell}$ i.i.d. 
with mean 0, variance 1 and finite moments of all orders;
\item[(A.iii$'$)] $\Sigma_p$ is a (possibly random) nonnegative
definite $p\times p$ matrix such that its ESD $F^{\Sigma_p}$ converges
almost surely in distribution to a probability distribution $H$ on
$[0,\infty)$
as $p\rightarrow\infty$; moreover, $H$ has a finite second moment;
\end{longlist}

\vspace*{-\baselineskip}

\renewcommand\thelonglist{(A.v)}
\renewcommand\labellonglist{\thelonglist}
\begin{longlist}[(A.iii$'$\hspace*{1pt})]
\item\label{asmconvw} the weights $w^n_\ell,  1\leq\ell\leq n,
n=1,2,\ldots,$ are all positive, and there exists $\kappa<\infty$
such that
the rescaled weights $(nw^n_\ell)$ satisfy
\[
\max_n \max_{\ell=1,\ldots,n}  (nw^n_\ell) \leq\kappa;
\]
moreover, almost surely, there exists a process $w_s\in D([0,1];\zz{R}_+)$
such that
%
\begin{equation}\label{eqconvintw}
\lim_n \sum_{1\leq\ell\leq n} \int_{(\ell- 1)/n}^{\ell/n}
|nw^n_\ell- w_s| \,ds = 0;
\end{equation}
\end{longlist}

\vspace*{-\baselineskip}

\renewcommand\thelonglist{(A.vi)}
\renewcommand\labellonglist{\thelonglist}
\begin{longlist}[(A.iii$'$\hspace*{1pt})]
\item\label{asmwdep} there exists a sequence $\eta_p = o(p)$ and a
sequence of index sets $\sI_p$ satisfying
$\sI_p\subset\{1,\ldots,p\}$ and $\# \sI_p \leq\eta_p$ such that
for all $n$ and all $\ell$, $w^n_\ell$ may depend
on $\mathbf{Z}_\ell^{(p)}$ but only on
$\{Z_\ell^{(p,j)}\dvtx j\in\sI_p\}$;
\end{longlist}

\vspace*{-\baselineskip}

\renewcommand\thelonglist{(A.vii)}
\renewcommand\labellonglist{\thelonglist}
\begin{longlist}[(A.iii$'$)]
\item\label{asmSigmanormrcv}
there exist $C<\infty$ and
$\delta<1/6$
such that for all $p$,
$\|\Sigma_p\|\leq Cp^{\delta}$ almost surely.
\end{longlist}
Define $S_p=\sum_{\ell=1}^n w^n_\ell\cdot\Sigma_p^{1/2} \mathbf{Z}^{(p)}_\ell(\mathbf{Z}^{(p)}_\ell)^T \Sigma_p^{1/2}$. Then, almost
surely, the ESD of $S_p$ converges in distribution to a probability
distribution $F^w$, which is
determined by $H$ and $(w_s)$ in that its Stieltjes transform
$m_{F^w}(z)$
is given by
%
\begin{equation}\label{eqSTFHWeighted}
m_{F^w}(z) = -\frac{1}{z}\int_{\tau\in\zz{R}} \frac{1}{\tau M(z)
+ 1}  \,dH(\tau),
\end{equation}
where $M(z)$, together with another function $\wt{m}(z)$, uniquely
solve the following equation
in $\zz{C}_+\times\zz{C}_+$:
%
\begin{equation}\label{eqMwtm}
\cases{\displaystyle M(z) = -\frac{1}{z} \int_0^1 \frac{w_s}{1+y \wt{m}(z) w_s}
\,ds,\vspace*{2pt}\cr
\displaystyle \wt{m}(z) = - \frac{1}{z} \int_{\tau\in\zz{R}}\frac{\tau}{\tau
M(z) + 1} \,d{H}(\tau).}
\end{equation}
\end{theorem}
%
\begin{remark}
Assumption (A.i$'$) can undoubtedly be weakened, for example, by using
the truncation and centralization technique
as in \citet{SB95} and \citet{Silverstein95}; or, a closer look at the
proof of Theorem \ref{thmMPWeightedCov} indicates that as long as
$Z^{(p,j)}_{\ell}$ has finite moments up to order
$k> 6/(1-6\delta)$,
the theorem is true and can be proved by exactly the same argument.
\end{remark}
%
\begin{remark}\label{rmkthm2}
If $w^n_\ell\equiv1/n$, then $w_s\equiv1$,\vspace*{-1pt}
and Theorem \ref{thmMPWeightedCov} reduces to Proposition~\ref
{propMPgen}. Moreover, if $w_s$ is not constant, that is, $w_s\not\equiv
\int_0^1 w_t \,dt$ on $[0,1)$, then except in the trivial case when $H$
is a delta measure at $0$,
the LSD $F^w\neq F$, where $F$ is the LSD in Proposition \ref{propMPgen}
determined by $H(\cdot/{\int_0^1 w_t \,dt})$.
See the supplementary article [\citet{ZL10supp}] for more details.
\end{remark}

Theorem \ref{thmMPWeightedCov} is proved in Section\vadjust{\goodbreak}
\ref{secpfMPw}.

A direct consequence of this theorem and Lemma
\ref{lemmadriftnegligible} below is the following
Mar\v{c}enko--Pastur type result for RCV matrices for
diffusion processes in class $\sC$.
We note that, thanks to Lemma \ref{lemmadriftnegligible} below (see
the remark after the proof of Lemma~\ref{lemmadriftnegligible} for
more explanations), regarding the drift process,
except requiring them to be uniformly bounded, we put no additional
assumption on them: they can be, for example, stochastic,
c\`{a}dl\`{a}g and dependant with each other. Furthermore, we allow
for dependence between the covolatility process and the
underlying Brownian motion---in other words, we allow for the
leverage effect. In the special case when $\gamma_t^{(p)}$ does not
change in $p$, is nonrandom and bounded,
and the observation times are equally spaced,
the (rather technical) assumptions (B.iii) and (B.iv) below are
trivially satisfied.\looseness=1

\begin{prop}\label{propMPRCV}
Suppose that for all $p$,
$\mathbf{X}^{(p)}_t$ is a $p$-dimensional process in class
$\mathcal{C}$ for some drift process
$\mu_t^{(p)}=(\mu_t^{(p,1)},\ldots,\mu_t^{(p,p)})^T$, covolatility
process $\Theta_t^{(p)} = \gamma^{(p)}_t \Lambda^{(p)}$ and
$p$-dimensional Brownian motion $\mathbf{W}_t^{(p)}=(
W_t^{(p,1)},\ldots,\allowbreak W_t^{(p,p)})^T$. Suppose further that:
\begin{longlist}[(B.iii)]
\item[(B.i)] 
there exists $C_0<\infty$ such that for all $p$ and all $j=1,\ldots
,p$, $|\mu_t^{(p,j)}|\leq C_0$
for all $t\in[0,1)$ almost surely;
\item[(B.ii)] $\breve{\Sigma}_p = \Lambda^{(p)} (\Lambda^{(p)})^T$
satisfies assumption \textup{(A.iii$'$)} and \textup{(A.vii)}
in Theorem~\ref{thmMPWeightedCov};
\item[(B.iii)] there exists a sequence $\eta_p = o(p)$ and a sequence of index
sets $\sI_p$ satisfying
$\sI_p\subset\{1,\ldots,p\}$ and $\# \sI_p \leq\eta_p$ such that
$\gamma^{(p)}_t$ may depend on $\mathbf{W}_t^{(p)}$ but only on
$\{W_t^{(p,j)}\dvtx j\in\sI_p\}$; moreover, there exists $C_1<\infty$
such that for all $p$, $|\gamma^{(p)}_t|\leq C_1$ for all
$t\in[0,1)$ almost surely; additionally, almost surely, there exists
$(\gamma_t) \in D([0,1];\zz{R})$ such that
\[
\lim_p \int_0^1 \bigl|\gamma^{(p)}_t - \gamma_t\bigr|  \,dt = 0;
\]
\item[(B.iv)] the observation times $\tau_{n,\ell}$ are independent of $\mathbf{X}_t$; moreover,
there exists $\kappa<\infty$ such that the observation durations $
\Delta\tau_{n,\ell}:=\tau_{n,\ell} - \tau_{n,\ell-1}$ satisfy
\[
\max_n \max_{\ell=1,\ldots,n} (n\cdot\Delta\tau_{n,\ell}) \leq
\kappa;
\]
additionally, almost surely, there exists a process $\upsilon_s\in
C([0,1);\zz{R}_+)$ such that
\[
\tau_{n,[ns]} \rightarrow
\Upsilon_s:=\int_0^s \upsilon_r\, dr
\qquad\mbox{as } n\rightarrow\infty\mbox{ for all } 0\leq s\leq1,
\]
where for any $x$, $[x]$ stands for its integer part.
\end{longlist}
Then, as $p\rightarrow\infty$, $F^{\Sigma_p^{\mathrm{RCV}}}$ converges almost surely
to a probability distribution~$F^w$ as specified in Theorem \ref{thmMPWeightedCov} for
$w_s = (\gamma_{\Upsilon_s})^2 \upsilon_s$.
\end{prop}

Proposition \ref{propMPRCV} demonstrates explicitly how the LSD of
RCV matrix
depends on the time-variability of the covolatility process. Hence, the
RCV matrix by itself cannot be used to make robust inference for the
ESD $F^{\Sigma_p}$ of the ICV matrix. If $(\gamma_s)$ [and hence $w_s
= (\gamma_{\Upsilon_s})^2 \upsilon_s$] is
known, then in principle, the equations (\ref{eqSTFHWeighted}) and
(\ref{eqMwtm}) can be used to recover $F^{\Sigma_p}$. However, in
general, $(\gamma_s)$ is unknown and
estimating the process $(\gamma_s)$ can be challenging and will
bring in more complication in the inference. Moreover, the equations~(\ref{eqSTFHWeighted}) and (\ref{eqMwtm}) are different from and
more complicated than the classical
Mar\v{c}enko--Pastur equation (\ref{eqSTFH0}), and in order to
recover $F^{\Sigma_p}$ based on these equations, one has to extend
existing algorithms [\citet{Karoui08}, \citet{Mestre08} and \citet
{BCY10} etc.] which are designed for~(\ref{eqSTFH0}).
Developing such an algorithm is of course of great interest, but we
shall not pursue this in the present article. We shall instead propose
an alternative estimator which overcomes these difficulties.

\subsection{Time-variation adjusted realized covariance (TVARCV) matrix}

$\!\!\!$Sup\-pose that a diffusion process $\mathbf{X}_t$
belongs to class $\mathcal{C}$. We define the \textit{time-variation
adjusted realized
covariance} (TVARCV) matrix as follows:
%
\begin{equation}\label{eqTVARCV}
\wh{\Sigma}_p:=\frac{\tr(\Sigma^{\mathrm{RCV}}_p)}{n}\cdot\sum_{\ell
=1}^n \frac{\Delta\mathbf{X}_\ell(\Delta
\mathbf{X}_\ell)^T}{|\Delta\mathbf{X}_\ell|^2}=\frac{\tr(\Sigma^{\mathrm{RCV}}_p)}{p} \wt{\Sigma}_p,
\end{equation}
where
for any vector $\mathbf{v}$, $|\mathbf{v}|$
stands for its Euclidean norm, and
%
\begin{equation}\label{eqsigmawt}
\wt{\Sigma}_p:=\frac{p}{n}\cdot\sum_{\ell=1}^n \frac{\Delta\mathbf{X}_\ell(\Delta\mathbf{X}_\ell)^T}{|\Delta\mathbf{X}_\ell|^2}.
\end{equation}
Let us first explain $\wt{\Sigma}_p$. Consider the simplest case
when $\mu_t\equiv0$, $\gamma_t$ deterministic, $\Lambda_t\equiv
I_{p\times p}$, and $\tau_{n,\ell}=\ell/n, \ell=0,1,\ldots, n$.
In this case, $\Delta\mathbf{X}_\ell=
\sqrt{\int_{(\ell-1)/n}^{\ell/n} \gamma_t^2  \,dt}\cdot\mathbf{Z}_\ell/\sqrt{n} $ where $\mathbf{Z}_\ell=(Z_\ell^{(1)},\ldots,Z_\ell
^{(p)})^T$ and $Z_\ell^{(j)}$'s are
i.i.d. standard normal.\vspace*{1pt} Hence, $\Delta\mathbf{X}_\ell(\Delta
\mathbf{X}_\ell)^T/|\Delta\mathbf{X}_\ell|^2= \mathbf{Z}_\ell(\mathbf{Z}_\ell
)^{T}/|\mathbf{Z}_\ell|^2$.
However, as $p\rightarrow\infty$, $|\mathbf{Z}_\ell|^2\sim p$, hence $\wt
{\Sigma}_p \sim1/n\cdot
\sum_{\ell=1}^n \mathbf{Z}_\ell(\mathbf{Z}_\ell)^{T}$, the latter
being the usual sample\vspace*{1pt} covariance matrix.
We will show that, first, $\tr(\Sigma^{\mathrm{RCV}}_p) \sim
\tr(\Sigma_p)$; and second, if $\mathbf{X}_t$ belongs to
class~$\mathcal{C}$ and satisfies certain additional assumptions, then
the LSD of $\wt{\Sigma}_p$ is related to that of
$\breve{\Sigma}_p$ via the Mar\v{c}enko--Pastur equation (\ref
{eqSTFH0}), where
%
\begin{equation}\label{eqSigmabr}
\breve{\Sigma}_p = \frac{p}{\tr(\Sigma_p)}{\Sigma_p}=\Lambda
\Lambda^T.
\end{equation}
Hence, the LSD of $ \wh{\Sigma}_p $ is also related to that of
$\Sigma_p$ via the same Mar\v{c}enko--Pastur equation.

We now state our assumptions.
Observe that about the drift process, again, except requiring them to
be uniformly bounded, we put no additional
assumption.\vadjust{\goodbreak}
Furthermore, we allow for the dependence between the covolatility
process and the
underlying Brownian motion, namely, the leverage effect.\vspace*{8pt}

\textit{Assumptions}:
{\renewcommand\thelonglist{(C.\roman{longlist})}
\renewcommand\labellonglist{\thelonglist}
\begin{longlist}[(C.viii)]
\item\label{asmdrift}
there exists $C_0<\infty$ such that for all $p$ and all $j=1,\ldots
,p$, $|\mu_t^{(p,j)}|\leq C_0$
for all $t\in[0,1)$ almost surely;
\item\label{asmgammadep}
there exist constants $C_1<\infty, 0\leq\delta_1 <1/2$, a sequence
$\eta_p<C_1 p^{\delta_1}$ and a sequence of index sets $\sI_p$ satisfying
$\sI_p\subset\{1,\ldots,p\}$ and $\# \sI_p \leq\eta_p$
such that $\gamma_t^{(p)}$ may depend on $\mathbf{W}_t^{(p)}$ but only on
$\{W_t^{(p,j)}\dvtx j\in\sI_p\}$;
moreover, there exists $C_2<\infty$ such that for all $p$, $|\gamma
_t^{(p)}|\in(1/C_2,C_2)$ for all
$t\in[0,1)$ almost surely;
\item\label{asmsigmabd}
there exists $C_3<\infty$ such that for all $p$ and for all $j$,
the individual volatilities $\sigma^{(j)}_t = \sqrt{(\gamma
_t^{(p)})^2\cdot\sum_{k=1}^p (\Lambda^{(p)}_{jk})^2} \in
(1/C_3,C_3)$ for all $t\in[0,1]$ almost surely;
\item\label{asmtrSigmagrow}
$\lim_{p\rightarrow\infty} \tr(\Sigma_p)/p $ $(\mbox{$=$}\lim_{p\rightarrow\infty} \int
_0^1(\gamma_t^{(p)})^2 \,dt): =\theta>0$ almost surely;
\item\label{asmSigmaLSD} almost surely, as $p\rightarrow\infty$, the ESD
$ F^{\Sigma_p}$
converges to a probability distribution $H$ on $[0,\infty)$;
\item\label{asmSigmanorm}
there exist $C_5<\infty$ and $0\leq\delta_2<1/2$ such that for all $p$,
$\|\Sigma_p\|\leq C_5p^{\delta_2}$ almost surely;
\item\label{asm2delta} the $\delta_1$ in \ref{asmgammadep}
and $\delta_2$ in \ref{asmSigmanorm}
satisfy that
$\delta_1 + \delta_2 < 1/2;$
\item\label{asmpn}
$p/n\rightarrow y\in(0,\infty)$ as $p\rightarrow\infty$;   and
\item\label{asmobstime} there exists $C_4<\infty$ such that for
all $n$,
\[
\max_{1\leq\ell\leq n} n \cdot(\tau_{n,\ell}-\tau_{n,\ell
-1})\leq C_4\qquad\mbox{almost surely};
\]
moreover, $\tau_{n,\ell}$'s are independent of $\mathbf{X}_t$.
\end{longlist}}

\noindent We have the following convergence theorem regarding the ESD of
our proposed estimator TVARCV matrix $\wh{\Sigma}_p$.
%
\begin{theorem}\label{thmconvedf}
Suppose that for all $p$, $\mathbf{X}_t =\mathbf{X}^{(p)}_t$
is a $p$-dimensional process in class $\mathcal{C}$
for some drift process $\mu_t^{(p)}=(\mu_t^{(p,1)},\ldots,\mu
_t^{(p,p)})^T$, covolatility process $\Theta_t^{(p)} = \gamma^{(p)}_t
\Lambda^{(p)}$ and $p$-dimensional Brownian motion
$\mathbf{W}_t^{(p)}$,
which satisfy assumptions
\ref{asmdrift}$\sim$\ref{asm2delta} above. Suppose also that
$p$ and $n$ satisfy~\ref{asmpn}, and the observation times satisfy
\ref{asmobstime}. Let
$\wh{\Sigma}_p$ be as in (\ref{eqTVARCV}). Then, as $p\rightarrow\infty$,
$F^{\wh{\Sigma}_p}$~converges almost surely to a probability
distribution $F$, which is
determined by $H$ through Stieltjes transforms via the same Mar\v
{c}enko--Pastur equation (\ref{eqSTFH0})
as in Proposition \ref{propMPgen}.
\end{theorem}

The proof of Theorem \ref{thmconvedf} is given in Section
\ref{secpfthm}.

The LSD $H$ of the targeting ICV matrix is in general not the same
as the LSD~$F$, but can be recovered from $F$ based on equation
(\ref{eqSTFH0}). In practice, when one has only finite number of
samples, the articles [\citet{Karoui08}, \citet{Mestre08} and \citet
{BCY10} etc.]
studied the estimation of the population spectral distribution based on
the sample covariance matrices.
In particular,
applying Theorem 2 of \citet{Karoui08} to our case yields.
%
\begin{cor}\label{coresdrec}
Let $H_p = F^{\wh{\Sigma}_p}$, and define $\wh{H}_p$ as in Theorem 2
of El~Ka\-roui (\citeyear{Karoui08}).
If $\|\Sigma_p\|$ are bounded in $p$, then, as $p\rightarrow\infty$,
$ \wh{H}_p \rightarrow H$ almost~surely.
\end{cor}

Therefore, when the dimension $p$ is large, based on the ESD of TVARCV
matrix~$\wh{\Sigma}_p$, we can estimate the spectrum of underlying
ICV matrix $\Sigma_p$ well.

\section{Proofs}\label{secproofs}

\subsection{Preliminaries}
\label{ssecprelim}

We collect some either elementary or well-known facts in the following. The
proofs are given in the supplemental article [\citet{ZL10supp}].
%
\begin{lemma}\label{lemmadriftnegligible}
Suppose that for each $p$, $\mathbf{v}_{\ell}^{(p)}=(v_\ell
^{(p,1)},\ldots,v_\ell^{(p,p)})^T$ and $\mathbf{w}_\ell^{(p)}=(w_\ell
^{(p,1)},\ldots,w_\ell^{(p,p)}),  \ell= 1,\ldots,n$, are all
$p$-dimensional vectors. Define
\[
\wt{S}_n =\sum_{\ell= 1}^n \bigl(\mathbf{v}_\ell^{(p)} + \mathbf{w}_\ell
^{(p)}\bigr)\cdot\bigl(\mathbf{v}_\ell^{(p)} + \mathbf{w}_\ell^{(p)}\bigr)^T
\quad\mbox{and}\quad
S_n = \sum_{\ell= 1}^n \mathbf{w}_\ell^{(p)} \bigl(\mathbf{w}_\ell^{(p)}\bigr)^T.
\]
If the following conditions are satisfied:
\begin{longlist}
\item$n=n(p)$ with $\lim_{p\rightarrow\infty} p/n = y>0$;
\item
there exists a sequence $\vep_p=o(1/\sqrt{p})$ such that for all $p$
and all $\ell$, all the entries of $\mathbf{v}_{\ell}^{(p)}$ are bounded
by $\vep_p$ in absolute value;
\item$\limsup_{p\rightarrow\infty} \tr(S_n)/p<\infty$ almost surely.
\end{longlist}
Then $L(F^{\wt{S}_n}, F^{S_n})\rightarrow0$ almost surely, where for any two
probability distribution functions $F$ and $G$, $L(F,G)$ denotes the
Levy distance between them.
\end{lemma}
%
%
\begin{lemma}[{[Lemma 2.6 of \citet{SB95}]}]\label{lemmatrdiff} Let
$z\in\zz{C}$ with
$v=\im(z)> 0$, $A$ and $B$ be $p\times p$ with $B$ Hermitian, and
$\mathbf{q}\in\zz{C}^p$. Then
\[
\bigl|\tr\bigl(\bigl((B-zI)^{-1} - (B+\tau\mathbf{q} \mathbf{q}^* -
zI)^{-1}\bigr)\cdot
A\bigr)\bigr|\leq\|A\|/v\qquad  \mbox{for all } \tau\in\zz{R}.
\]
\end{lemma}

The following two lemmas are similar to Lemma 2.3 in
\citet{Silverstein95}.

\begin{lemma}\label{lemmanormsum} Let $w\in\zz{C}$ with $\re
(w)\geq0$, and
$A$ be an Hermitian nonnegative definite matrix. Then
$ \|(wA + I)^{-1}\|\leq1$.
\end{lemma}
%
\begin{lemma}\label{lemmadiffest} Let $w_1,w_2\in\zz{C}$ with $\re
(w_1)\geq0$ and $\re(w_2)\geq
0$, $A$ be a $p\times p$ Hermitian nonnegative definite matrix,
$B$\vadjust{\goodbreak}
any $p\times p$ matrix, and $\mathbf{q}\in\zz{C}^p$. Then:
\begin{longlist}
\item
$
|\tr( B( (w_1 A + I)^{-1} - (w_2 A +
I)^{-1}))|
\leq p\cdot|w_1 - w_2|\cdot\|B\|\cdot\|A\|;
$
\item
$
|\mathbf{q}^* B(w_1 A+ I)^{-1}\mathbf{q} - \mathbf{q}^* B(w_2 A+ I)^{-1}\mathbf{q}|
\leq|w_1 - w_2|\cdot|\mathbf{q}|^2 \|B\|\cdot\|A\|.
$
\end{longlist}
\end{lemma}
%
\begin{lemma}\label{lemmanormdiff}
For any Hermitian matrix $A$ and $z\in\zz{C}$ with $\im(z)=v>0$,
$ \|(A-zI)^{-1}\|\leq1/v$.
\end{lemma}

Both Lemmas \ref{lemmanormsum} and \ref{lemmadiffest} require
the real part of $w$ (or $w_1$, $w_2$) to be nonnegative. In our proof
of Theorem
\ref{thmMPWeightedCov}, the requirements will be fulfilled thanks to
the following lemma.
%
\begin{lemma}\label{lemmarealpos}
Let $z=iv\in\zz{C}$ with $v>0$, $A$ be a $p\times p$ Hermitian
nonnegative definite matrix, $\mathbf{q}\in\zz{C}^p$, $\tau>0$.
Then
\[
-\frac{1}{z}\cdot\frac{1}{1+\tau\mathbf{q}^* (A-zI)^{-1}
\mathbf{q}}
\in Q_1=\{z\in\zz{C}\dvtx \re(z)\geq0,\im(z) \geq0\}.
\]
\end{lemma}
%
\begin{lemma}\label{lemmapostrace}
Let $z=iv\in\zz{C}$ with $v>0$, $A$ be any $p\times p$ matrix, and
$B$ be a~$p\times p$ Hermitian
nonnegative definite matrix.
Then
\mbox{$ \tr(A (B - zI)^{-1} A^*) \in Q_1$}.
\end{lemma}

\begin{lemma}\label{lemmaunitm}
Suppose that $w_s\in D([0,1);\zz{R}_+)$. Then for any $y\in\zz{C}$,
the equation
\[
\int_0^1 \frac{1}{1 + z w_s} \,ds = y
\]
admits at most one solution in $Q_1$.
\end{lemma}

The following result is an immediate consequence of Lemma 2.7 of \citet{BS98}.
%
\begin{lemma}\label{lemmaconcstdnormal}
For $\mathbf{X}=(X^{(1)},\ldots,X^{(p)})^T$ where $X^{(j)}$'s are i.i.d.
random variables such that $EX^{(1)}=0, E|X^{(1)}|^2 = 1$, and
$E|X^{(1)}|^{2k}<\infty$ for some $2\leq k\in\zz{N}$,
there exists $C_k\geq0$, depending only on $k$, $E|X^{(1)}|^4$ and
$E|X^{(1)}|^{2k}$,
such that for any $p\times p$ nonrandom matrix $A$,
\[
E|\mathbf{X}^* A \mathbf{X} - \tr(A)|^{2k}\leq C_k(\tr(AA^*))^{k}\leq C_k
p^k\|A\|^{2k}.
\]
\end{lemma}

\begin{prop}[{[Theorem 2 of \citet{Geronimo03}]}]\label{propstieltjesconv}
Suppose\break that~$P_n$ are real probability measures with
Stieltjes transforms $m_n(z)$. Let $K\subseteq\zz{C}_+$ be an
infinite set with a limit point in $\zz{C}_+$. If $\lim
m_n(z):=m(z)$ exists for all $z\in K$, then there exists a
probability measure $P$ with Stieljes transform~$m(z)$ if and only if
%
\begin{equation}\label{eqconditionst}
\lim_{v\rightarrow\infty} iv\cdot m(iv) = -1,
\end{equation}
in which case $P_n\rightarrow P$ in distribution.
\end{prop}

\subsection{\texorpdfstring{Proof of Proposition \protect\ref{propnotconv}}{Proof of Proposition 3}}
\label{seccountereg}

By assumption, $\gamma_t$ is positive and non-constant on $[0,1)$, and
is c\`{a}dl\`{a}g, in particular, right-continuous; moreover, \mbox{$\int
_0^1\gamma_t^2 \,dt = \sigma^2$}. Hence, there exists
$\delta>0$ and $[c,d]\subseteq[0,1]$ such that
\[
\gamma_t \geq\sigma(1+\delta)\qquad\mbox{for all } t\in[c,d].
\]
Therefore, if $[(\ell-1)/n,\ell/n]\subseteq[c,d]$,
\[
\Delta\mathbf{X}_{\ell} (\Delta\mathbf{X}_{\ell})^T
\eqd\int_{(\ell-1)/n}^{\ell/n}\gamma_t^2 \,dt\cdot\mathbf{Z}_\ell(\mathbf{Z}_\ell)^T
\geq\frac{(1+\delta)^2}{n}\cdot\sigma^2 \mathbf{Z}_\ell(\mathbf{Z}_\ell)^T,
\]
where
$\mathbf{Z}_\ell=(Z_\ell^{(1)},\ldots,Z_\ell^{(p)})^T$ consists of independent
standard normals.
Hence, if we let
$ J_n=\{\ell\dvtx [(\ell-1)/n,\ell/n]\subseteq[c,d]\}$
and
\[
\Gamma_p=\sum_{\ell\in J_n} \Delta\mathbf{X}_{\ell}
(\Delta\mathbf{X}_{\ell})^T,\qquad
 \Lambda_p =\frac{\sigma^2}{(n(d-c))}\cdot\sum_{\ell\in
J_n}\mathbf{Z}_\ell(\mathbf{Z}_\ell)^T,
\]
then for any $x\geq0$, by Weyl's Monotonicity theorem [see, e.g.,
Corollary~4.3.3 in \citet{HornJohnson}],
%
\begin{equation}\label{eqSpecMono}
F^{\Sigma^{\mathrm{RCV}}_p}(x) \leq F^{\Gamma_p}(x)
\leq F^{\Lambda_p}\bigl(x/[(1+\delta)^2(d-c)]\bigr).
\end{equation}
Now note that $\#J_n \sim(d-c)n$, hence if $p/n\rightarrow y$, by
Proposition \ref{propMP}, $F^{\Lambda_p}$ will converge almost
surely to the Mar\v{c}enko--Pastur law with ratio index
$y'=y/(d-c)$ and scale index $\sigma^2$, which has density on
$[a(y'),b(y')]$ with functions $a(\cdot)$ and $b(\cdot)$ defined
by (\ref{eqMPSupp}). By the formula of $b(\cdot)$,
\[
(1+\delta)^2(d-c) b(y')
=(1+\delta)\cdot\sigma^2 (1+\delta)\bigl(y + 2\sqrt{(d-c)y} +
d-c\bigr).
\]
Hence, for any $\vep>0$, there exists $y_c>0$ such that for all
$y\geq y_c$,
\[
(1+\delta)^2(d-c) b(y')
\geq(1+\delta)\cdot\sigma^2\bigl(\bigl(1+\sqrt{y}\bigr)^2+\vep\bigr)
=(1+\delta)\bigl(b(y) + \sigma^2\vep\bigr),
\]
that is,
\[
\frac{(b(y) + \sigma^2\vep)}{(1+\delta)^2(d-c)}
\leq\frac{b(y')}{1+\delta}.
\]
By (\ref{eqSpecMono}), when the above inequality holds,
\[
\limsup F^{\Sigma^{\mathrm{RCV}}_p}\bigl(b(y) + \sigma^2\vep\bigr)
\leq\mbox{MP}^{(y',\sigma^2)}\bigl(b(y')/(1+\delta)\bigr)<1.
\]
%

\subsection{\texorpdfstring{Proof of Theorem \protect\ref{thmMPWeightedCov}}{Proof of Theorem 1}}
\label{secpfMPw}

To prove Theorem \ref{thmMPWeightedCov}, following the strategies in
\citet{MP67}, \citet{Silverstein95}, \citet{SB95}, we
will work with Stieltjes transforms.
\begin{pf*}{Proof of Theorem \ref{thmMPWeightedCov}}
For notational ease, we shall sometimes omit the sub/superscripts $p$
and $n$ in the arguments below: thus, we write
$\mathbf{Z}_\ell$ instead of
$\mathbf{Z}^{(p)}_\ell$, $w_\ell$ instead of $w^n_\ell$,
$\Sigma$ instead of $\Sigma_p$, $S$ instead of $S_p$, etc. Also
recall that $y_n =
p/n$, which
converges to $y>0$.

By assumption (A.vi) we may, without loss of generality,
assume that the weights $w_\ell$ are independent of $\mathbf{Z}_\ell$'s.
This is because, if we let $\wt{\mathbf{Z}}_\ell$ be the result of replacing
$Z^{(p,j)}_\ell, j\in\sI_p, $
with independent random variables with the same distribution that are
also independent of $w_\ell$, and $\wt{S}\,{:=}\,\sum_{\ell=1}^n w_\ell
\,{\cdot}\allowbreak\Sigma^{1/2} \wt{\mathbf{Z}}_\ell(\wt{\mathbf{Z}}_\ell)^T \Sigma
^{1/2}$, then $\mbox{rank}(\wt{S}-S)\leq2\eta_p$, and so
by the rank inequality
\[
\|F^{A} - F^{B}\|\leq\frac{\mbox{rank}(A-B)}{p}\qquad\mbox{for any }
A, B\ p\times p \mbox{ symmetric matrices}
\]
[see, e.g., Lemma 2.2 in \citet{Bai99}], $\wt{S}$ and $S$ must
have the same LSD.

We proceed according to whether $H$ is a delta measure at~$0$ or not.
If~$H$ is a delta measure at $0$, we claim that $F^w$ is also a delta
measure at~$0$, and the conclusion of the theorem holds. The reason is
as follows. By assump\-tion~\ref{asmconvw},
\[
S=\sum_{\ell=1}^n w_\ell\cdot\Sigma^{1/2} \mathbf{Z}_\ell
(\mathbf{Z}_\ell)^T \Sigma^{1/2}
\leq\frac{\kappa}{n} \sum_{\ell=1}^n \Sigma^{1/2} \mathbf{Z}_\ell
(\mathbf{Z}_\ell)^T \Sigma^{1/2}:=\kappa\ol{S}.
\]
Hence by Weyl's Monotonicity theorem again, for any $x\geq0$
\[
F^{S}(x)\geq F^{\ol{S}}(x/\kappa).
\]
However, it follows easily from Proposition \ref{propMPgen} that $
F^{\ol{S}}$ converges to the delta measure at $0$, hence so does $F^{S}$.

Below we assume that $H$ is \textit{not} a delta measure at $0$.

Let $I= I_{p\times p}$ be the $p\times p$ identity matrix, and
\[
m_n:=m_n(z) = \frac{\tr((S - z I)^{-1})}{p}
\]
be the Stieltjes transform of $F^{S}$.
By Proposition \ref{propstieltjesconv},
in order to show that~$F^{S}$
converges, it suffices to prove that for all $z=iv$ with $v>0$
sufficiently large,
$\lim_n m_n(z):=m(z)$ exists, and that $m(z)$ satisfies condition
(\ref{eqconditionst}).

We first show the convergence of $m_n(z)$ for $z=iv$ with $v>0$
sufficiently large.
Since for all $n$, $|m_n(z)|\leq1/v$, it suffices to show that
$\{m_n(z)\}$ has at most one limit.

For notational ease, we denote by $\mathbf{r}_\ell=\Sigma^{1/2} \mathbf{Z}_\ell$.
We first show that
%
\begin{equation}\label{eqrinorm}
\max_{\ell=1,\ldots,n} \bigl||\mathbf{r}_\ell|^2/p - h\bigr|
= {\max_{\ell=1,\ldots,n}} |\mathbf{Z}_\ell^T \Sigma\mathbf{Z}_\ell/p -
h| \rightarrow0  \qquad\mbox{almost surely,}\hspace*{-34pt}
\end{equation}
where $h = \int_0^\infty x \,d H(x)$.
In fact, by Lemma \ref{lemmaconcstdnormal} and
assumptions~(A.i$'$)\break
and~\ref{asmSigmanormrcv}, for any
$k\in\zz{N}$,
\[
E|\mathbf{Z}_\ell^T \Sigma\mathbf{Z}_\ell- \tr(\Sigma)|^{2k}\leq C_kp^k
p^{2\delta
k}\qquad \mbox{for all } 1\leq\ell\leq n.
\]
Using Markov's inequality we
get that for any $\vep>0$,
\[
P\bigl(|\mathbf{Z}_\ell^T \Sigma\mathbf{Z}_\ell- \tr(\Sigma)|\geq p \vep\bigr)
\leq\frac{C_kp^k p^{2\delta k}}{p^{2k}\vep^{2k}}
=\frac{C_k \vep^{-2k}}{p^{(1-2\delta)k}}
\qquad \mbox{for all } 1\leq\ell\leq n.
\]
Hence, choosing
$k> 2/(1-2\delta)$,
using Borel--Cantelli and that $n=O(p)$ yield
%
\begin{equation}\label{eqrinorm0}
\max_{\ell=1,\ldots,n} |\mathbf{Z}_\ell^T \Sigma\mathbf{Z}_\ell/p -
\tr(\Sigma)/p| \rightarrow0
\qquad \mbox{almost surely.}
\end{equation}
The convergence (\ref{eqrinorm}) follows.

Next, let
%
\begin{equation}\label{eqMn}
M_n=M_n(z) = - \frac{1}{z}\sum_{\ell=1}^n \frac{w_\ell}{1+w_\ell
\mathbf{r}_\ell^T (S_{(\ell)} - zI)^{-1}
\mathbf{r}_\ell},
\end{equation}
where
\[
S_{(\ell)}: = \sum_{j\neq\ell} w_j \mathbf{r}_j \mathbf{r}_j^T
= S - w_\ell\mathbf{r}_\ell\mathbf{r}_\ell^T.
\]
Note that by Lemma \ref{lemmarealpos}, for any $\ell$,
%
\begin{equation}\label{eqwipos}
- \frac{1}{z}\frac{1}{1+ w_\ell\mathbf{r}_\ell^T (S_{(\ell)} - zI)^{-1}
\mathbf{r}_\ell}\in Q_1.
\end{equation}
We shall show that
%
\begin{equation}\label{eq2Sttrsame}
\frac{1}{p}\tr(-z M_n \Sigma- zI)^{-1} -m_n \rightarrow0
\qquad \mbox{almost surely}.
\end{equation}

Observe the following identity: for any $p\times p$ matrix $B$,
$\mathbf{q}\in\zz{R}^p$ and $\tau\in\zz{C}$ for which $B$ and
$B+\tau\mathbf{q} \mathbf{q}^T$ are both invertible,
%
\begin{equation}\label{eqmultidecom}
\mathbf{q}^T (B+ \tau\mathbf{q} \mathbf{q}^T)^{-1} = \frac{1}{1+\tau\mathbf{q}^T
B^{-1} \mathbf{q}} \mathbf{q}^T
B^{-1};
\end{equation}
see equation (2.2) in \citet{SB95}.
Writing
\[
S - zI - (-z M_n \Sigma- zI)
=\sum_{\ell=1}^n w_\ell\mathbf{r}_\ell\mathbf{r}_\ell^T - (-zM_n \Sigma),
\]
taking the inverse, using (\ref{eqmultidecom}) and the definition
(\ref{eqMn}) of $M_n$ yield
\begin{eqnarray*}
&&(-zM_n \Sigma- zI)^{-1} - (S - zI)^{-1}\\
&&\qquad=(-zM_n \Sigma- zI)^{-1}\Biggl(\sum_{\ell=1}^n w_\ell\mathbf{r}_\ell
\mathbf{r}_\ell^T -
(-zM_n \Sigma)\Biggr)(S - zI)^{-1}\\
&&\qquad=-\frac{1}{z}\sum_{\ell=1}^n\frac{ w_\ell}{1+ w_\ell\mathbf{r}_\ell
^T (S_{(\ell)} - zI)^{-1}
\mathbf{r}_\ell}(M_n \Sigma+ I)^{-1} \mathbf{r}_\ell\mathbf{r}_\ell
^T\bigl(S_{(\ell)} -
zI\bigr)^{-1}\\
&&\qquad\quad{} + \frac{1}{z}\sum_{\ell=1}^n \frac{w_\ell}{1+ w_\ell\mathbf{r}_\ell^T (S_{(\ell)} - zI)^{-1}
\mathbf{r}_\ell}(M_n \Sigma+ I)^{-1} \Sigma(S -
zI)^{-1}.
\end{eqnarray*}
Taking trace and dividing by $p$ we get
\[
\frac{1}{p}\tr(-zM_n S - zI)^{-1} -m_n
=\frac{1}{z}\sum_{\ell=1}^n\frac{w_\ell}{1+w_\ell\mathbf{r}_\ell^T
(S_{(\ell)} - zI)^{-1}
\mathbf{r}_\ell}\cdot d_\ell,
\]
where
\[
d_\ell= \frac{1}{p}\bigl(\tr\bigl((M_n \Sigma+ I)^{-1} \Sigma(S -
zI)^{-1}\bigr) - \mathbf{r}_\ell^T\bigl(S_{(\ell)} -
zI\bigr)^{-1}(M_n \Sigma+ I)^{-1} \mathbf{r}_\ell\bigr).
\]
By
(5.2) in the proof of Lemma \ref{lemmarealpos} in the supplementary
article [\citet{ZL10supp}],
$\re(\mathbf{r}_\ell^T (S_{(\ell)} - zI)^{-1} \mathbf{r}_\ell))\geq0$.
Hence,
%
\begin{equation}\label{eqbddcoef}
\biggl|\frac{1}{1+ w_\ell\mathbf{r}_\ell^T (S_{(\ell)} - zI)^{-1}
\mathbf{r}_\ell}\biggr|\leq1.
\end{equation}
Therefore in order to show (\ref{eq2Sttrsame}), by assumption
\ref{asmconvw}, it suffices to
prove
%
\begin{equation}\label{eqdito0}
{\max_{\ell=1,\ldots,n}} |d_\ell| \rightarrow0  \qquad\mbox{almost surely}.
\end{equation}

Define
\[
M_{(\ell)}= M_{(\ell)}(z)
= - \frac{1}{z}\sum_{j\neq\ell} \frac{w_j}{1+w_j \mathbf{r}_j^T
(S_{(j,\ell)} - zI)^{-1}
\mathbf{r}_j},
\]
where $ S_{(j,\ell)}: = \sum_{i\neq j, \ell} w_i \mathbf{r}_i \mathbf{r}_i^T
= S - (w_j \mathbf{r}_j \mathbf{r}_j^T+ w_\ell\mathbf{r}_\ell\mathbf{r}_\ell
^T)$. Observe that
for every~$\ell$, $M_{(\ell)}$ is independent of $\mathbf{Z}_\ell$.
%
\begin{claim}
For any $z=iv$ with $v>0$ and any $\vep<1/2-\delta$,
%
\begin{equation}\label{eqdiffM}
\max_{\ell=1,\ldots, n} p^\vep\bigl|M_{(\ell)}(z) - M_n(z)\bigr| \rightarrow0
\qquad\mbox{almost surely.}
\end{equation}
\end{claim}
\begin{pf}
Define
\[
\wt{m}_n(z)=\frac{\tr( {\Sigma}^{1/2} (S - zI)^{-1}{\Sigma
}^{1/2})}{p},
\]
which belongs to $Q_1$ by Lemma \ref{lemmapostrace}, and
%
\begin{equation}\label{dfnMtilde}
\cases{\displaystyle \wt{M}_{n}= \wt{M}_{n}(z)
= - \frac{1}{z}\sum_{j=1}^n \frac{w_j}{1+ y_n n w_j \wt{m}_n(z)},\vspace*{2pt}\cr
\displaystyle \wt{M}_{(\ell)}= \wt{M}_{(\ell)}(z)
= - \frac{1}{z}\sum_{j\neq\ell} \frac{w_j}{1+ y_n n w_j \wt{m}_n(z)}\vspace*{2pt}\cr
\hphantom{\wt{M}_{(\ell)}}=\displaystyle \wt{M}_n(z) + \frac{1}{z} \frac{w_\ell}{1+ y_n n w_\ell\wt{m}_n(z)}.}
\end{equation}
Then by a similar argument for (\ref{eqbddcoef}) and using
assumption \ref{asmconvw},
\[
\max_{\ell=1,\ldots,n} \bigl|\wt{M}_{(\ell)}(z) - \wt{M}_n(z)\bigr|\leq
\frac{\kappa}{nv}.\vadjust{\goodbreak}
\]
Hence, it suffices to show that
%
\begin{eqnarray}\label{eqMMtilde}
p^\vep|M_{n}(z) - \wt{M}_n(z)| &\rightarrow& 0\quad
\mbox{and }\nonumber\\[-8pt]\\[-8pt]
\max_{\ell=1,\ldots, n} p^\vep\bigl|M_{(\ell)}(z) - \wt{M}_{(\ell
)}(z)\bigr| &\rightarrow& 0\qquad
\mbox{almost surely.}\nonumber
\end{eqnarray}
We shall only prove the second convergence. In fact,
\begin{eqnarray*}
&&M_{(\ell)}(z) - \wt{M}_{(\ell)}(z)\\
&&\qquad= - \frac{1}{z}\sum_{j\neq\ell} \frac{w_j\cdot y_n n w_j }{(1+w_j
\mathbf{r}_j^T (S_{(j,\ell)} - zI)^{-1} \mathbf{r}_j)\cdot(1+ y_n n w_j \wt
{m}_n(z))} \cdot \zeta_{j,\ell},
\end{eqnarray*}
where
\[
\zeta_{j,\ell} = \wt{m}_n(z) - \frac{\mathbf{r}_j^T (S_{(j,\ell)} - zI)^{-1}
\mathbf{r}_j}{p}.
\]
Since for all $j$,
\[
\biggl|\frac{w_j\cdot y_n n w_j }{(1+w_j \mathbf{r}_j^T (S_{(j,\ell)} -
zI)^{-1} \mathbf{r}_j)\cdot(1+ y_n n w_j \wt{m}_n(z))}\biggr|
\leq\frac{\kappa^2 y_n}{n},
\]
it suffices to show that
%
\begin{equation}\label{eqzeta}
\max_{\ell=1,\ldots, n} \max_{j\neq\ell} p^\vep| \zeta_{j,\ell
} | \rightarrow0\qquad\mbox{almost surely.}
\end{equation}
To prove this, recall that $\mathbf{r}_\ell= \Sigma^{1/2} \mathbf{Z}_\ell
$, by Lemma \ref{lemmaconcstdnormal} and the independence between
$\mathbf{Z}_\ell$ and $ \Sigma^{1/2}
(S_{(j,\ell)} - zI)^{-1}{\Sigma}^{1/2}$, for any $k\in\zz{N}$,
%
\begin{eqnarray}\label{eqdifftracenorm0}
&&
E\bigl|\mathbf{Z}_\ell^T {\Sigma}^{1/2} \bigl(S_{(j,\ell)} -
zI\bigr)^{-1}{\Sigma}^{1/2}
\mathbf{Z}_\ell- \tr\bigl( {\Sigma}^{1/2} \bigl(S_{(j,\ell)} -
zI\bigr)^{-1}{\Sigma}^{1/2}\bigr) \bigr|^{2k}\nonumber\\
&&\qquad\leq C_k p^k \cdot E\bigl(\bigl\|{\Sigma}^{1/2} \bigl(S_{(j,\ell
)} - zI\bigr)^{-1}{\Sigma}^{1/2}\bigr\|^{2k}\bigr)\\
&&\qquad\leq \frac{C_k p^k \cdot E\|{\Sigma}\|
^{2k} }{v^{2k}}
\leq\frac{C p^k p^{2\delta k}}{v^{2k}},\nonumber
\end{eqnarray}
where in the last line we used Lemma
\ref{lemmanormdiff} and assumption \ref{asmSigmanormrcv}.
Hence, for any $\vep< 1/2 - \delta$, choosing
$k> 3/(1-2\delta-2\vep)$
and
using Borel--Cantelli again, we get
%
\begin{eqnarray}\label{eqconcnum0}
&&
\max_{\ell=1,\ldots,n} \max_{j\neq\ell}
p^\vep\biggl| \frac{\mathbf{Z}_\ell^T
{\Sigma}^{1/2} (S_{(j,\ell)} - zI)^{-1}{\Sigma}^{1/2}
\mathbf{Z}_\ell}{p} \nonumber\\[-8pt]\\[-8pt]
&&\hspace*{44pt}\qquad{}  - \frac{\tr( {\Sigma}^{1/2} (S_{(j,\ell)} -
zI)^{-1}{\Sigma}^{1/2})}{p}\biggr| \rightarrow0.\nonumber
\end{eqnarray}
Furthermore, by Lemma \ref{lemmatrdiff} and assumption \ref
{asmSigmanormrcv},
recall that $\wt{m}_n(z)=
\tr( {\Sigma}^{1/2} (S - zI)^{-1}{\Sigma}^{1/2})/p$,
%
\begin{eqnarray}\label{eqdifftrace0}
&&\max_{\ell=1,\ldots,n}\max_{j\neq\ell}
\biggl|\frac{1}{p}\tr\bigl( {\Sigma}^{1/2} \bigl(S_{(j,\ell)} -
zI\bigr)^{-1}{\Sigma}^{1/2}\bigr)- \wt{m}_n(z) \biggr|\nonumber\\[-8pt]\\[-8pt]
&&\qquad\leq 2 \frac{ \|{\Sigma}\| }{p v}
\leq\frac{C p^\delta}{pv}.
\nonumber
\end{eqnarray}
The convergence (\ref{eqzeta}) follows.
\end{pf}

We now continue the proof of the theorem.
Recall that $\mathbf{r}_\ell= \Sigma^{1/2}\mathbf{Z}_\ell$, and~$\mathbf{Z}_\ell$ consists of i.i.d. random variables with finite
moments of all orders.
By Lem\-ma~\ref{lemmadiffest}(ii) and (\ref{eqwipos}),
%
\begin{eqnarray}\label{eqdiffnorm}
&&
{\max_{\ell=1,\ldots,n}} \frac{|\mathbf{r}_\ell^T(S_{(\ell)}\!-\!
zI)^{-1}(M_n \Sigma\!+\!I)^{-1} \mathbf{r}_\ell\!-\!\mathbf{r}_\ell^T(S_{(\ell)}\!-\!zI)^{-1}(M_{(\ell)}
\Sigma\!+\!I)^{-1}\mathbf{r}_\ell|}{p}\hspace*{-25pt}\nonumber\\
&&\quad\leq{\max_{\ell=1,\ldots,n}} \frac{ |M_{(\ell)}\!-\!M_n(z)|\cdot
|\mathbf{r}_\ell|^2 \cdot\|(S_{(\ell)}\!-\!zI)^{-1}\|\cdot\| \Sigma\|}{p}\\
&&\quad\leq{\max_{\ell=1,\ldots,n}} \frac{|M_{(\ell)}\!-\!M_n(z)| \cdot C
p^\delta}{v }\cdot\frac{|\mathbf{r_\ell}|^2}{p}\!\rightarrow\!0,
\nonumber
\end{eqnarray}
where in the last line we used Lemma \ref{lemmanormdiff}, assumption
\ref{asmSigmanormrcv}, the assumption that $\delta<1/6$ (and
hence $\delta< 1/2-\delta$) and (\ref{eqdiffM}), and
(\ref{eqrinorm}).

Furthermore, similar to (\ref{eqdifftracenorm0}), by Lemma \ref
{lemmaconcstdnormal} and the
independence between $\mathbf{Z}_\ell$ and $ \Sigma^{1/2}
(S_{(\ell)} - zI)^{-1}(M_{(\ell)} \Sigma+
I)^{-1}{\Sigma}^{1/2}$, for any $k\in\zz{N}$,
\begin{eqnarray*}
&&
E\bigl|\mathbf{Z}_\ell^T {\Sigma}^{1/2} \bigl(S_{(\ell)} -
zI\bigr)^{-1}\bigl(M_{(\ell)} {\Sigma} + I\bigr)^{-1}{\Sigma}^{1/2}
\mathbf{Z}_\ell\\
&&\quad{} - \tr\bigl( {\Sigma}^{1/2} \bigl(S_{(\ell)} -
zI\bigr)^{-1}\bigl(M_{(\ell)}
{\Sigma} + I\bigr)^{-1}{\Sigma}^{1/2}\bigr) \bigr|^{2k}\\
&&\qquad\leq C_k p^k \cdot E\bigl(\bigl\|{\Sigma}^{1/2} \bigl(S_{(\ell)}
- zI\bigr)^{-1}\bigl(M_{(\ell)}
{\Sigma} + I\bigr)^{-1}{\Sigma}^{1/2}\bigr\|^{2k}\bigr)\\
&&\qquad\leq \frac{C_k p^k \cdot E\|{\Sigma}\|^{2k} }{v^{2k}}
\leq\frac{C p^k p^{2\delta k}}{v^{2k}},
\end{eqnarray*}
where in the last line we use Lemmas
\ref{lemmanormdiff}, \ref{lemmanormsum} and (\ref{eqwipos}),
and assumption \ref{asmSigmanormrcv}. Hence, choosing
$k> 2/(1-2\delta)$
and
using Borel--Cantelli again, we get
%
\begin{eqnarray}\label{eqconcnum}
&&
\max_{\ell=1,\ldots,n}
\biggl| \frac{\mathbf{Z}_\ell^T
{\Sigma}^{1/2} (S_{(\ell)} - zI)^{-1}(M_{(\ell)}
{\Sigma} + I)^{-1}{\Sigma}^{1/2}
\mathbf{Z}_\ell}{p} \nonumber\\[-8pt]\\[-8pt]
&&\hspace*{23.2pt}\quad{} - \frac{\tr( {\Sigma}^{1/2} (S_{(\ell)} -
zI)^{-1}(M_{(\ell)}
{\Sigma} + I)^{-1}{\Sigma}^{1/2})}{p}\biggr| \rightarrow
0.
\nonumber
\end{eqnarray}
Furthermore, by Lemmas \ref{lemmadiffest}(i),
\ref{lemmanormsum} and (\ref{eqwipos}), the assumption that
$\delta<1/6$ (and hence $2\delta< 1/2-\delta$) and (\ref
{eqdiffM}), and assumption \ref{asmSigmanormrcv},
%
\begin{eqnarray}\label{eqdifftrace}
&&\max_{\ell=1,\ldots,n}
\biggl|\frac{1}{p}\tr\bigl( {\Sigma}^{1/2} \bigl(S_{(\ell)} -
zI\bigr)^{-1}\bigl(M_{(\ell)}
{\Sigma} + I\bigr)^{-1}{\Sigma}^{1/2}\bigr)\nonumber\\
&&\qquad\hspace*{12pt}{} - \frac{1}{p}\tr\bigl( {\Sigma}^{1/2} \bigl(S_{(\ell)} -
zI\bigr)^{-1}\bigl(M_n
{\Sigma} + I\bigr)^{-1}{\Sigma}^{1/2}\bigr)\biggr|\nonumber\\[-8pt]\\[-8pt]
&&\qquad\leq \max_{\ell=1,\ldots,n} \bigl|M_{(\ell)} - M_n\bigr|\cdot\bigl\|\bigl(S_{(\ell
)} - zI\bigr)^{-1} \bigr\|\cdot\|{\Sigma}\|^2\nonumber\\
&&\qquad\leq \max_{\ell=1,\ldots,n} \bigl|M_{(\ell)} - M_n\bigr|\cdot\frac{C
p^{2\delta}}{v} \rightarrow0.
\nonumber
\end{eqnarray}

Finally, similar to (\ref{eqdifftrace0}), by Lemmas \ref
{lemmatrdiff} and \ref{lemmanormsum}, and assumption \ref
{asmSigmanormrcv},
%
\begin{eqnarray}\label{eqdifftrace2}
&&
\max_{\ell=1,\ldots,n}
\biggl|\frac{1}{p}\tr\bigl( {\Sigma}^{1/2} \bigl(S_{(\ell)} -
zI\bigr)^{-1}(M_n {\Sigma} + I)^{-1}{\Sigma}^{1/2}\bigr)\nonumber\\
&&\qquad\hspace*{11pt}{}  - \frac{1}{p}\tr\bigl( {\Sigma}^{1/2} (S -
zI)^{-1}(M_n{\Sigma} + I)^{-1}{\Sigma}^{1/2}\bigr)\biggr|\\
&&\qquad\leq \frac{ \|(M_n{\Sigma} + I)^{-1} \|\cdot\|{\Sigma}\| }{p v}
\leq\frac{C p^\delta}{pv} \rightarrow0.
\nonumber
\end{eqnarray}

Combining
(\ref{eqdiffnorm}),
(\ref{eqconcnum}), (\ref{eqdifftrace}) and (\ref
{eqdifftrace2}), we see that
(\ref{eqdito0}), and hen\-ce~(\ref{eq2Sttrsame}) holds.

Now we are ready to show that $\{m_n(z)\}$ admits at most one
limit.
%
\begin{claim}\label{claimconvM} Suppose that $m_{n_k}(z)$ converges
to $m(z)$, then
%
\begin{equation}\label{eqnM}
M_{n_k}(z)\rightarrow M(z):= -\frac{1}{z} \int_0^1 \frac{w_s}{1+y \wt{m}(z)
w_s} \,ds
\neq0,
\end{equation}
where $\wt{m}(z)$ is the unique solution in $Q_1=\{z\in\zz{C}\dvtx \re
(z)\geq0,\im(z) \geq0\}$ to the following equation:
%
\begin{equation}\label{eqntm}
\int_0^1 \frac{1}{1 + y \wt{m}(z) w_s} \,ds = 1- y\bigl( 1 + z m(z) \bigr).
\end{equation}
\end{claim}
\begin{pf}
Writing
\[
S - z I + z I = \sum_{\ell=1}^n w_\ell\mathbf{r}_\ell\mathbf{r}_\ell^T,
\]
right-multiplying both sides by $(S - z I)^{-1}$ and using (\ref
{eqmultidecom}) we
get
\[
I + z (S - z I)^{-1} =\sum_{\ell=1}^n \frac{w_\ell\mathbf{r}_\ell
\mathbf{r}_\ell^T(S_{(\ell)} - zI)^{-1}}{1+ w_\ell\mathbf{r}_\ell^T
(S_{(\ell)} - zI)^{-1} \mathbf{r}_\ell}.
\]
Taking trace and dividing by $n$ we get
\[
y_n + z y_n m_n(z) = 1 - \frac{1}{n}\sum_{\ell=1}^n \frac
{1}{1+w_\ell\mathbf{r}_\ell^T (S_{(\ell)} - zI)^{-1}
\mathbf{r}_\ell},
\]
where, recall that, $m_n(z) = \tr((S - z I)^{-1})/p$
is the Stieltjes transform of $F^{S}$.
Hence, if $m_{n_k}(z)\rightarrow m(z)$, then
%
\begin{eqnarray}\label{eqconvtm}
&&\frac{1}{n_k}\sum_{\ell=1}^{n_k}
\frac{1}{1+y_{n_k} n_k w_\ell \cdot \mathbf{r}_\ell^T
(S_{(\ell)} - zI)^{-1}\mathbf{r}_\ell/p_k}\nonumber\\
&&\qquad=\frac{1}{n_k}\sum_{\ell=1}^{n_k}
\frac{1}{1+w_\ell \mathbf{r}_\ell^T (S_{(\ell)} - zI)^{-1}
\mathbf{r}_\ell}\\
&&\qquad=1-y_{n_k}\bigl(1+z m_{n_k}(z)\bigr)
\to 1-y\bigl( 1 + z m(z) \bigr).\nonumber
\end{eqnarray}
However, by the same arguments for (\ref{eqconcnum0}) and (\ref
{eqdifftrace0}) we have
%
\begin{equation}\label{eqconvtm1}
\max_{\ell=1,\ldots,n}
\biggl|\frac{\mathbf{r}_\ell^T
(S_{(\ell)} - zI)^{-1}
\mathbf{r}_\ell}{p} - \frac{\tr( {\Sigma}^{1/2} (S_{(\ell)} -
zI)^{-1}{\Sigma}^{1/2})}{p}\biggr| \rightarrow 0
\end{equation}
and
%
\begin{equation}\label{eqconvtm2}
\max_{\ell=1,\ldots,n}
\biggl|
\frac{\tr( {\Sigma}^{1/2} (S_{(\ell)} - zI)^{-1}{\Sigma
}^{1/2})}{p}
-
\wt{m}_n(z)\biggr|
\rightarrow0,
\end{equation}
where, recall that
$
\wt{m}_n(z)=\tr( {\Sigma}^{1/2} (S - zI)^{-1}{\Sigma
}^{1/2})/p,
$
which belongs to $Q_1$ by Lemma \ref{lemmapostrace}. Then by (\ref
{eqconvtm}), assumption \ref{asmconvw}
and Lemma \ref{lemmaunitm}, $\wt{m}_{n_k}(z)$ must also converge,
and the limit, denoted by $\wt{m}(z)\in Q_1$,
must be the unique solution in $Q_1$ to the equation (\ref{eqntm}).
Now by
(\ref{dfnMtilde}), (\ref{eqMMtilde})
and assumption~\ref{asmconvw},
we get the convergence for $M_{n_k}(z)$ in the claim. That $M(z)\neq0$
follows from the expression and that $\wt{m}(z) \in Q_1$.
\end{pf}

We now continue the proof of the theorem. By the
convergence of $F^{{\Sigma}_p}$ to~${H}$ and the previous claim,
\[
\frac{\tr((-zM_{n_k}(z) {\Sigma} - zI)^{-1})}{p} \rightarrow
-\frac{1}{z}\int_{\tau\in\zz{R}} \frac{1}{\tau M(z) + 1 }
\,d{H}(\tau).
\]
But (\ref{eq2Sttrsame}) implies that
%
\begin{equation}\label{eqnm}
m(z) =-\frac{1}{z} \int_{\tau\in\zz{R}} \frac{1}{\tau M(z) + 1}
\,d{H}(\tau).
\end{equation}

Observing that $M(z)\neq0$, $\re(M(z))\geq0$, and $H$ is not a delta
measure at $0$, we obtain that $|m(z)|<1/|z|$. Hence $1+zm(z) \neq0$,
and by (\ref{eqntm}), $\wt{m}(z)\neq0$. Based on this, we can
get\vadjust{\goodbreak}
another expression for $M(z)$, as follows. By (\ref{eqnM}), we have
%
\begin{eqnarray}\label{eqnMf2}
M(z)&=& -\frac{1}{z} \int_0^1 \frac{w_s}{1+y \wt{m}(z) w_s}
\,ds\nonumber\\[-2pt]
&=&-\frac{1}{z}\cdot\frac{1}{y \wt{m}(z)}\cdot\biggl(1- \int_0^1
\frac{1}{1+y \wt{m}(z) w_s} \,ds\biggr)\nonumber\\[-9.5pt]\\[-9.5pt]
&=&-\frac{1}{z}\cdot\frac{1}{y \wt{m}(z)}\cdot\bigl(1- \bigl(1-y\bigl( 1 + z
m(z) \bigr)\bigr)\bigr)\nonumber\\[-2pt]
&=&-\frac{1}{z}\cdot\frac{1 + zm(z)}{\wt{m}(z)},
\nonumber\vspace*{-1pt}
\end{eqnarray}
where in the third line we used the definition (\ref{eqntm}) of $\wt{m}(z)$.

We can then derive another formula for $\wt{m}(z)$.
By (\ref{eqnm}),
\[
1+ zm(z) = 1 -\int_{\tau\in\zz{R}} \frac{1}{\tau M(z) + 1}
\,d{H}(\tau)
= M(z) \int_{\tau\in\zz{R}}\frac{\tau}{\tau M(z) + 1} \,d{H}(\tau)\vspace*{-1pt}
\]
by using that $H$ is a probability distribution. Dividing both sides
by\break
$-z \wt{m}(z)(\mbox{$\neq$} 0)$
and using (\ref{eqnMf2}) yield
\[
M(z)=-\frac{(1+ zm(z))}{z\wt{m}(z)}
= -\frac{M(z)\int_{\tau\in\zz{R}}{\tau}/({\tau M(z) + 1})
\,d{H}(\tau)}{z \wt{m}(z)},\vspace*{-1pt}
\]
and hence since $M(z)\neq0$,
%
\begin{equation}\label{eqntmf2}
\wt{m}(z) = -\frac{1}{z} \int_{\tau\in\zz{R}}\frac{\tau}{\tau
M(z) + 1} \,d{H}(\tau).\vspace*{-1pt}
\end{equation}

Observe that by Lemma \ref{lemmapostrace} and (\ref{eqwipos}),
for any $n$, both $m_n(z)$ and
$M_n(z)$ belong to~$Q_1$, hence so do $m(z)$ and $M(z)$. We proceed to
show that for those $z=iv$ with $v$ sufficiently large, there is at
most one triple $(m(z),M(z),\allowbreak\wt{m}(z))\in Q_1\times Q_1\times Q_1$
that solves the equations (\ref{eqnm}), (\ref{eqnM}) and (\ref
{eqntmf2}). In fact, if there are two different triples
$(m_i(z),M_i(z), \wt{m}_i(z)), i=1,2$, both satisfying (\ref
{eqnm}), (\ref{eqnM}) and (\ref{eqntmf2}). Then necessarily,
$M_1(z)\neq M_2(z)$ and $\wt{m}_1(z)\neq\wt{m}_2(z)$.
Now by (\ref{eqnM}),
\[
M_1(z) - M_2(z) = -\frac{1}{z}\cdot y \bigl(\wt{m}_2(z) - \wt{m}_1(z)\bigr)
\int_0^1 \frac{w_s^2}{(1+y \wt{m}_1(z) w_s)(1+y \wt{m}_2(z) w_s)} \,ds\vspace*{-1pt}
\]
by (\ref{eqntmf2}),
\[
\wt{m}_1(z) - \wt{m}_2(z) = -\frac{1}{z} \bigl(M_2(z) - M_1(z)\bigr)
\int_{\tau\in\zz{R}}\frac{\tau^2}{(\tau M_1(z) + 1)(\tau M_2(z) +
1)} \,d{H}(\tau).\vspace*{-1pt}
\]
Therefore,
%
\begin{eqnarray}\label{eq1}
1 &=& \frac{y}{z^2} \int_0^1 \frac{w_s^2}{(1+y \wt{m}_1(z) w_s)(1+y
\wt{m}_2(z) w_s)} \,ds\nonumber\\[-9.5pt]\\[-9.5pt]
&&{}\times \int_{\tau\in\zz{R}}\frac{\tau^2}{(\tau M_1(z) + 1)(\tau
M_2(z) + 1)} \,d{H}(\tau).
\nonumber\vspace*{-1pt}\vadjust{\goodbreak}
\end{eqnarray}
However, since $(M_i(z),\wt{m}_i(z))\in Q_1\times Q_1$, $i=1,2$,
\[
\biggl|\int_0^1 \frac{w_s^2}{(1+y \wt{m}_1(z) w_s)(1+y \wt{m}_2(z)
w_s)} \,ds\biggr|
\leq\int_0^1 w_s^2  \,ds<\infty
\]
and
\[
\biggl|\int_{\tau\in\zz{R}}\frac{\tau^2}{(\tau M_1(z) + 1)(\tau
M_2(z) + 1)} \,d{H}(\tau)\biggr|
\leq\int_{\tau\in\zz{R}}\tau^2  \,d{H}(\tau)<\infty.
\]
Hence, for $z=iv$ with $v$ sufficiently large, (\ref{eq1}) cannot be true.

It remains to verify (\ref{eqconditionst}), that is,
$
\lim_{v\rightarrow\infty} iv\cdot m(iv) = -1.
$
In fact, using~(\ref{eqnm}) we get that
%
\begin{equation}\label{eqmcond}
iv\cdot m(iv) = - \int_{\tau\in\zz{R}} \frac{1}{1+\tau M(iv)}
\,d{H}(\tau).
\end{equation}
Since $\re(M(iv))\geq0$, $ | 1/(1+\tau M(iv))|\leq1$
for all $\tau\geq0$.
Moreover, by (\ref{eqnM}) and that $\re(\wt{m}(z))\geq 0$,
$|M(iv)|\,{\leq}\,1/v\,{\cdot}\,\int_0^1 w_s\,ds$,
hence by the dominated~con\-vergence theorem, the right-hand side of
(\ref{eqmcond}) converges to
${-}1$ as $v\,{\rightarrow}\,\infty$.~%
\end{pf*}

\subsection{\texorpdfstring{Proof of Theorem \protect\ref{thmconvedf}}{Proof of Theorem 2}}\label{secpfthm}
The TVARCV matrix has the form of weighted sample covariance matrices
as studied in Theorem \ref{thmMPWeightedCov};
however, assumption~\ref{asmwdep} therein is not satisfied, and we
need another proof.

Theorem \ref{thmconvedf} is a direct consequence of the
following two convergence results.

\begin{prop}\label{propRVLDP} Under assumption \ref
{asmtrSigmagrow}, namely, suppose that
\[
\lim_{p\rightarrow\infty} \tr(\Sigma_p)/p = \theta,
\]
then, almost surely,
$
\lim_{p\rightarrow\infty} \tr(\Sigma^{\mathrm{RCV}}_p)/p= \theta.
$
\end{prop}

The proof is given in the supplemental article [\citet{ZL10supp}].

Next, recall that $\breve{\Sigma}_p$ and $\wt{\Sigma}_p$ are
defined by
(\ref{eqSigmabr}) and (\ref{eqsigmawt}), respectively.

\begin{prop}\label{propconvedf} $\!\!\!$Under the assumptions of Theorem
\ref{thmconvedf}, both $F^{\breve{\Sigma}_p}$ and~$F^{\wt{\Sigma}_p}$
converge almost surely. $F^{\breve{\Sigma}_p}$
converges to $\breve{H}$ defined by
%
\begin{equation}\label{eqbreveH}
\breve{H} (x) = H(\theta x) \qquad\mbox{for all } x\geq0.
\end{equation}
The LSD $\wt{F}$ of $\wt{\Sigma}_p$ is determined by
$\breve{H}$ in that its Stieltjes transform
$m_{\wt{F}}(z)$
satisfies the equation
\[
m_{\wt{F}}(z) = \int_{\tau\in\zz{R}} \frac{1}{\tau(1-y(1+z
m_{\wt{F}}(z))) -z}  \,d\breve{H}(\tau).
\]
\end{prop}

This can be proved in very much the same way as Theorem \ref
{thmMPWeightedCov}, by working with Stieltjes
transforms. However, a much simpler and transparent proof is as
follows.\vadjust{\goodbreak}
\begin{pf*}{Proof of Proposition \ref{propconvedf}}
The convergence of $F^{\breve{\Sigma}_p}$ is obvious since
\[
F^{\breve{\Sigma}_p}(x) = F^{\Sigma_p}\bigl(\tr(\Sigma_p)/p\cdot
x\bigr)\qquad\mbox{for all } x\geq0.
\]

We now show the convergence of $F^{\wt{\Sigma}_p}$. As in the proof
of Theorem \ref{thmMPWeightedCov}, for notational ease, we shall
sometimes omit the superscript $p$ in the arguments below: thus, we write
${\bolds\mu}_t$ instead of
${\bolds\mu}^{(p)}_t$, $\gamma_t$ instead of $\gamma^{(p)}_t$,
$\Lambda$ instead of $\Lambda^{(p)}$, etc.

First, note that
\[
\Delta\mathbf{X}_\ell=\int_{\tau_{n,\ell-1}}^{\tau_{n,\ell}} \mathbf{\mu}_t  \,dt + \Lambda\cdot\int_{\tau_{n,\ell-1}}^{\tau_{n,\ell
}} \gamma_t \,d\mathbf{W}_t
:= \sqrt{\int_{\tau_{n,\ell-1}}^{\tau_{n,\ell}} \gamma_t^2
\,dt}\cdot(\mathbf{v}_\ell+ \Lambda\cdot
\mathbf{Z}_\ell),
\]
where
\[
\mathbf{v}_\ell=\pmatrix{
v_\ell^{(1)}\cr \vdots\vspace*{2pt}\cr v_\ell^{(p)}}
=\frac{ \int_{\tau_{n,\ell-1}}^{\tau_{n,\ell}} {\bolds\mu}_t  \,dt}
{\sqrt{\int_{\tau_{n,\ell-1}}^{\tau_{n,\ell}} \gamma_t^2  \,dt}}
\quad\mbox{and}\quad
\mathbf{Z}_\ell=\pmatrix{
Z_\ell^{(1)}\cr \vdots\vspace*{2pt}\cr Z_\ell^{(p)}}
=\frac{\int_{\tau_{n,\ell-1}}^{\tau_{n,\ell}} \gamma_t \,d\mathbf{W}_t}
{\sqrt{\int_{\tau_{n,\ell-1}}^{\tau_{n,\ell}} \gamma_t^2  \,dt}}.
\]
By performing an orthogonal transformation if necessary, without loss
of generality, we may assume that the index set $\sI_p\subset\{
1,\ldots,\eta_p\}$.
Then by assumptions \ref{asmgammadep} and \ref{asmobstime},
for $j>\eta_p$, $Z_\ell^{(j)}$ are i.i.d. $N(0,1)$.
Write $\mathbf{U}_\ell=(Z_\ell^{(1)},\ldots,Z_\ell^{(\eta_p)})^T$ and
$\mathbf{D}_\ell=(Z_\ell^{(\eta_p+1)},\ldots,Z_\ell^{(p)})^T$.
With the above notation,~$\wt{\Sigma}_p$ can be rewritten as
%
\begin{equation}\label{eqwtSigmaneat}
\wt{\Sigma}_p = y_n \sum_{\ell=1}^n \frac{ \Delta\mathbf{X}_\ell
(\Delta
\mathbf{X}_\ell)^T}{|\Delta\mathbf{X}_\ell|^2}
= y_n\sum_{\ell=1}^n \frac{(\mathbf{v}_\ell+ \Lambda\mathbf{Z}_\ell)
(\mathbf{v}_\ell+ \Lambda\mathbf{Z}_\ell)^T}{|\mathbf{v}_\ell+ \Lambda\mathbf{Z}_\ell|^2}.
\end{equation}

By assumptions \ref{asmdrift}, \ref{asmgammadep} and \ref
{asmobstime}, there exists $C>0$ such that $|v_\ell^{(j)}|\leq
C/\sqrt{n}$ for all $j$ and $\ell$, hence $|\mathbf{v}_\ell|$'s are
uniformly bounded.
We will show that
%
\begin{eqnarray}\label{eqldpendo}
&&\max_{\ell=1,\ldots,n} \bigl||\Lambda\mathbf{Z}_\ell|^2/p - 1
\bigr|\nonumber\\[-8pt]\\[-8pt]
&&\qquad= \max_{\ell=1,\ldots,n} |\mathbf{Z}_\ell^T\breve{\Sigma}_p \mathbf{Z}_\ell/p - 1 |\rightarrow0
\qquad\mbox{almost surely},\nonumber
\end{eqnarray}
which clearly implies that
%
\begin{equation}\label{eqldpnorm}
\max_{\ell=1,\ldots,n} \bigl||\mathbf{v}_\ell+ \Lambda\mathbf{Z}_\ell|^2/p -
1 \bigr|
\rightarrow0 \qquad\mbox{almost surely}.
\end{equation}

To prove (\ref{eqldpendo}), write
\[
\breve{\Sigma}_p=\pmatrix{
A & B \cr
B^T & C},
\]
where $A, B$ and $C$ are $\eta_p\times\eta_p, \eta_p\times(n-\eta
_p)$ and $(n-\eta_p)\times(n-\eta_p)$ matrices, respectively. Then
\[
\mathbf{Z}_\ell^T\breve{\Sigma}_p \mathbf{Z}_\ell= \mathbf{U}_\ell^T A \mathbf{U}_\ell+ 2 \mathbf{D}_\ell^T B^T \mathbf{U}_\ell
+ \mathbf{D}_\ell^T C \mathbf{D}_\ell.
\]
By a well-known fact about the spectral norm,
\[
\|A\|\leq\|\breve{\Sigma}_p\|,\qquad
\|B\|\leq\|\breve{\Sigma}_p\|\quad
\mbox{and}\quad\|C\|\leq\|\breve{\Sigma}_p\|.
\]
In particular, by assumptions \ref{asmgammadep}, \ref
{asmSigmanorm} and \ref{asm2delta},
\[
0\leq\tr(A) \leq\eta_p \cdot\|\breve{\Sigma}_p\|\leq C p^{\delta
_1 + \delta_2}=o(p),
\]
hence $\tr(C)/p = (\tr(\breve{\Sigma}_p) - \tr(A))/p\rightarrow1$. Now
using the fact that $\mathbf{D}_\ell$ consists of i.i.d. standard
normals and by the same proof as that for (\ref{eqrinorm0}) we get
%
\begin{equation}\label{eqldpendo1}
\max_{\ell=1,\ldots,n} |\mathbf{D}_\ell^T C \mathbf{D}_\ell/p - 1 |\rightarrow0
\qquad\mbox{almost surely}.
\end{equation}
To complete the proof of (\ref{eqldpendo}), it then suffices to show that
\[
\max_{\ell=1,\ldots,n} \frac{|\mathbf{U}_\ell^T A \mathbf{U}_\ell
|}{p}\rightarrow0
\quad\mbox{and}\quad\max_{\ell=1,\ldots,n} \frac{|\mathbf{D}_\ell^T B^T
\mathbf{U}_\ell|}{p}\rightarrow0 \qquad\mbox{almost surely}.
\]
We shall only prove the first convergence; the second one can be proved
similarly. We have
%
\begin{equation}\label{eqUlnorm}
|\mathbf{U}_\ell^T A \mathbf{U}_\ell|\leq\|A\|\cdot|\mathbf{U}_\ell|^2\leq
C_5 p^{\delta_2}\cdot|\mathbf{U}_\ell|^2.
\end{equation}
Observe that for all $1\leq i\leq\eta_p$, by assumption \ref
{asmgammadep},
\[
\bigl|Z^{(i)}_\ell\bigr|^2 =\frac{|\int_{\tau_{n,\ell-1}}^{\tau_{n,\ell}}
\gamma_t \,d\mathbf{W}_t|^2}
{\int_{\tau_{n,\ell-1}}^{\tau_{n,\ell}} \gamma_t^2  \,dt}
\leq\frac{C_2^2}{\Delta\tau_{n,\ell}}\cdot\biggl|\int_{\tau
_{n,\ell-1}}^{\tau_{n,\ell}} \gamma_t \,d\mathbf{W}_t\biggr|^2.
\]
By the Burkholder--Davis--Gundy inequality, we then get that for any $k\in
\zz{N}$, there exists $\lambda_k>0$ such that
%
\begin{equation}\label{eqrtnmoments}
E\bigl|Z^{(i)}_\ell\bigr|^{2k}\leq\lambda_k C_2^{4k}.
\end{equation}
Now we are ready to show that ${\max_{\ell=1,\ldots,n}} |\mathbf{U}_\ell
^T A \mathbf{U}_\ell|/p\rightarrow0$. In fact, for any $\vep>0$, for any $k\in
\zz{N}$, by Markov's inequality, (\ref{eqUlnorm}), H\"{o}lder's
inequality and~(\ref{eqrtnmoments}),
\begin{eqnarray*}
P\Bigl( {\max_{\ell=1,\ldots,n}} |\mathbf{U}_\ell^T A \mathbf{U}_\ell|
\geq p \vep\Bigr)
&\leq& \sum_{\ell=1}^n P( |\mathbf{U}_\ell^T A \mathbf{U}_\ell| \geq
p \vep)\\
&\leq& \sum_{\ell=1}^n \frac{E|\mathbf{U}_\ell^T A \mathbf{U}_\ell|^{k}
}{p^{k} \vep^{k} }\\
&\leq& \sum_{\ell=1}^n \frac{C_5^k p^{k\delta_2}\cdot[ (\eta_p
\cdot\lambda_k C_2^{4k})\cdot\eta_p^{k-1} ]}{p^{k} \vep^{k}}\\
&\leq& C p^{1+k\delta_2 + k\delta_1 - k}.
\end{eqnarray*}
By assumption \ref{asm2delta}, $\delta_1 + \delta_2<1/2<1$,
hence by choosing $k$ to be large enough, the right hand side will be
summable in $p$, hence by Borel--Cantelli, almost surely, ${\max_{\ell
=1,\ldots,n}} |\mathbf{U}_\ell^T A \mathbf{U}_\ell|/p\rightarrow0$.

We now get back to $\wt{\Sigma}_p$ as in (\ref{eqwtSigmaneat}). By
(\ref{eqldpnorm}), for any $\vep>0$, almost surely, for all $ n$
sufficiently large, for all $\ell=1,\ldots,n$,
\[
p(1-\vep)\leq|\mathbf{v}_\ell+ \Lambda\mathbf{Z}_\ell|^2\leq p(1+\vep).
\]
Hence, almost surely, for all $n$ sufficiently large,
\[
\frac{1}{1+\vep} \wt{S}_p
\leq\wt{\Sigma}_p
= y_n\sum_{\ell=1}^n \frac{(\mathbf{v}_\ell+ \Lambda\mathbf{Z}_\ell)
(\mathbf{v}_\ell+ \Lambda\mathbf{Z}_\ell)^T}{|\mathbf{v}_\ell+ \Lambda\mathbf{Z}_\ell|^2}
\leq\frac{1}{1-\vep} \wt{S}_p,
\]
where $\wt{S}_p = 1/n\cdot\sum_{1\leq\ell\leq n}(\mathbf{v}_\ell+
\Lambda\mathbf{Z}_\ell) (\mathbf{v}_\ell+ \Lambda\mathbf{Z}_\ell)^T$.
Hence, by Weyl's Monotonicity theorem,
for any $x\geq0$,
%
\begin{equation}\label{eqFsandwich}
F^{\wt{S}_p}\bigl((1+\vep) x\bigr) \geq F^{\wt{\Sigma}_p}(x)
\geq F^{\wt{S}_p}\bigl((1-\vep) x\bigr).
\end{equation}

Next, by Lemma \ref{lemmadriftnegligible}, $\wt{S}_p$ has the same
LSD as
$S_p\,{:=}\,1/n\sum_{1\leq\ell\leq n} \Lambda\mathbf{Z}_\ell(\mathbf{Z}_\ell)^T \Lambda^T$. Moreover,
by using the same trick as in the beginning of the proof of
Theorem \ref{thmMPWeightedCov}, $F^{S_p}$ has the same limit as
$F^{S'_p}$, where $S'_p=1/n\sum_{1\leq\ell\leq n} \Lambda\wt{\mathbf{Z}}_\ell(\wt{\mathbf{Z}}_\ell)^T \Lambda^T$, and $\wt{\mathbf{Z}}_\ell$
consists of i.i.d. standard normals. For $F^{S'_p}$, it follows
easily from Proposition \ref{propMPgen} that it converges to $\wt
{F}$. Moreover, by Theorems 1.1 and 2.1 in \citet{SC95}, $\wt{F}$ is
differentiable and in particular continuous at all $x>0$. It follows
from (\ref{eqFsandwich}) that $F^{\wt{\Sigma}_p}$ must also
converge to $\wt{F}$.
\end{pf*}

\section{Simulation studies}\label{secsimulation}

In this section, we present some simulation studies to illustrate
the behavior of ESDs of RCV and TVARCV matrices.
In particular, we show that the ESDs of RCV matrices that have the
same targeting ICV matrix $\Sigma_p$ can be quite different from
each other, depending on the time variability of the covolatility
process. Our proposed estimator, the TVARCV matrix $\wh{\Sigma}_p$,
in contrast, has a very stable ESD.

We use in particular a reference curve which is the
Mar\u{c}enko--Pastur law. The reason we compare the ESDs of RCV and
TVARCV matrices with the Mar\u{c}enko--Pastur law is that
the Mar\u{c}enko--Pastur law is the LSD of $\Sigma_p^{\mathrm{RCV}^0}$ defined
in (\ref{eqRCV0}), which is the RCV matrix estimated from sample
paths of constant volatility that has the same targeting ICV matrix as
$\Sigma_p^{\mathrm{RCV}}$.
As we will see soon in the following two subsections,
when the covolatility process is time varying, the ESD of RCV matrix
can be very different from the Mar\u{c}enko--Pastur law, while the ESD
of TVARCV matrix always matches the Mar\u{c}enko--Pastur law
very well.

In the simulation below, we assume that $\Lambda=I$, or in other
words, $\mathbf{X}_t$
satisfies
(\ref{eqXsimplest})
with $\gamma_t$ a deterministic (scalar) process, and $\mathbf{W}_t$ a $p$-dimensional standard Brownian motion. The observation
times are taken to be equidistant:
$\tau_{n,\ell}=\ell/n, \ell=0,1,\ldots,n$.

We present simulation results of two different designs: one when
$\gamma_t$ is piecewise constant, the other when $\gamma_t$ is
continuous (and non-constant).
In both cases, we
compare the ESDs of the RCV and TVARCV matrices. Results
for different dimension $p$ and observation frequency $n$ are
reported.

In all the figures below, we use red solid lines to represent the
LSDs of~$\Sigma^{\mathrm{RCV}_0}$ given by the Mar\u{c}enko--Pastur law,
black dashed line to represent the ESDs of RCV matrices, blue
bold longdashed line to represent the ESDs of TVARCV matrices.

\subsection{Design \textup{I}, piecewise constants}

We first consider the case when the volatility path follows
piecewise constants. More specifically, we take $ \gamma_t$ to be
\[
\gamma_t = \cases{\sqrt{0.0007}, &\quad  $t\in
[0,1/4)\cup[3/4,1]$,\cr
\sqrt{0.0001}, &\quad $t\in[1/4,3/4)$.}
\]

\begin{figure}

\includegraphics{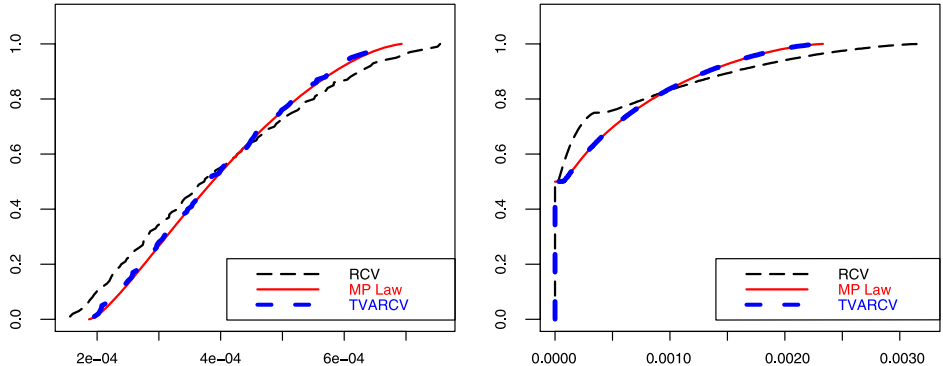}

\caption{Left panel: $p=100$, $n=1\mbox{,}000$; right panel: $p=2\mbox{,}000$,
$n=1\mbox{,}000$.}
\label{Figp100n1000dzn3}
\end{figure}

In Figure \ref{Figp100n1000dzn3}, we compare the ESDs of RCV and
TVARCV matrices for different pairs of $p$ and
$n$, with the LSD of $\Sigma^{\mathrm{RCV}_0}$ given by the
Mar\u{c}enko--Pastur law as reference.

%
%

%
\begin{figure}

\includegraphics{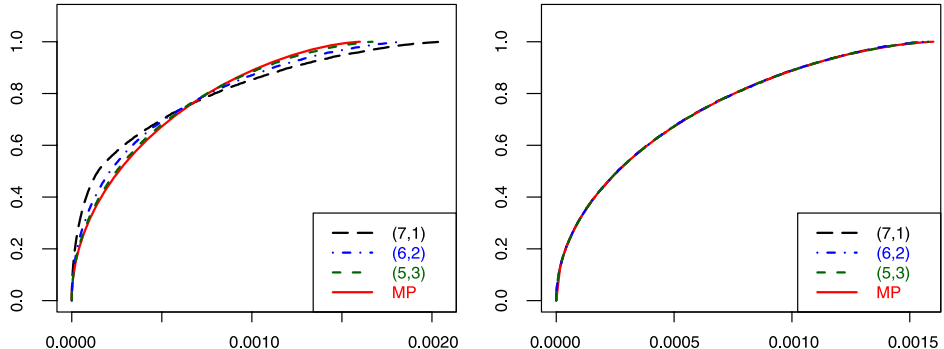}

\caption{Comparisons, different values of piecewise constants ($a, b$)
as shown in the legend, which
are such that the targeting ICV matrix is the same; the red solid curve
is the
Mar\u{c}enko--Pastur law (the LSD of $\Sigma^{\mathrm{RCV}^0}$). $p$ and $n$
are both taken to be $1\mbox{,}000$. Left panel: RCV; right panel: TVARCV.}
\label{FigcompRV}
\end{figure}

We see from Figure \ref{Figp100n1000dzn3} that:
\begin{itemize}
\item the ESDs of RCV matrices
are very different from the LSD given by the
Mar\u{c}enko--Pastur law (the LSD of $\Sigma^{\mathrm{RCV}^0}$);
\item the ESDs of TVARCV matrices follow the LSD given by the Mar\u
{c}enko--Pastur law very
well, for both pairs of $p$ and $n$, even when $p$ is small
compared with $n$.
\end{itemize}

In fact, the dependence of the ESD of RCV matrix on the time
variability of covolatility process can be seen more clearly from
Figure \ref{FigcompRV}, where we
consider the same design but different values for $\gamma_t$:
\[
\gamma_t = \cases{a^{1/2}\times10^{-2}, &\quad  $t\in
[0,1/4)\cup[3/4,1]$,\cr
b^{1/2}\times10^{-2}, &\quad  $t\in[1/4,3/4)$,}\qquad
\mbox{where } a + b = 8.
\]
We plot the ESDs of
RCV and TVARCV matrices for the case when
$p=n=1\mbox{,}000$, in the left and right panel, respectively.
The curves' corresponding parameters
$(a,b)$ are reported in the legend. Note that since all pairs of
$(a,b)$ have
the same summation,
in all cases the targeting ICV matrices are the same.

We see clearly from Figure \ref{FigcompRV}
that, the ESDs of RCV matrices can be very different
from each other even though the RCV matrices are estimating
the same ICV matrix; while for TVARCV matrices, the
ESDs are almost identical.

\subsection{Design \textup{II}, continuous paths}

We illustrate in this subsection the case when the volatility processes
have continuous sample paths. In
particular, we assume that $\mathbf{X}_t$ satisfies
(\ref{eqXsimplest}) with
\[
\gamma_t = \sqrt{0.0009 + 0.0008 \cos(2\pi t)},\qquad t\in[0,1].
\]

\begin{figure}

\includegraphics{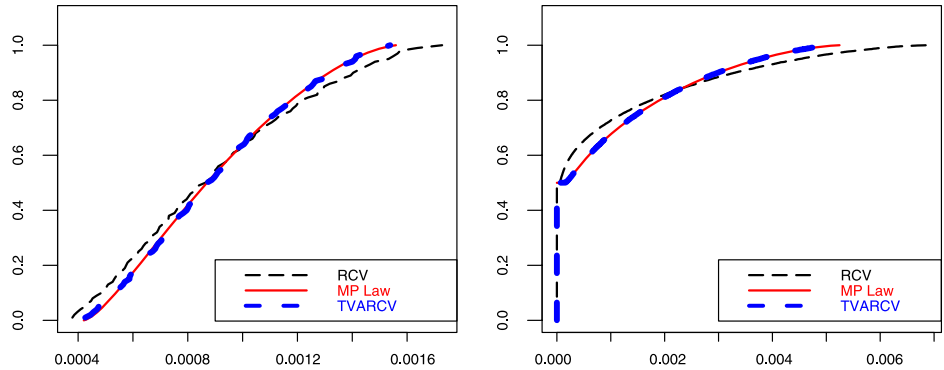}

\caption{Left panel: $p=100$, $n=1\mbox{,}000$; right panel: $p=2\mbox{,}000$, $n=1\mbox{,}000$.}
\label{Figp100n1000cos}
\vspace*{-4pt}
\end{figure}

We see from Figure \ref{Figp100n1000cos} similar phenomena as in Design I about the ESDs of RCV and
TVARCV matrices for different
pairs of $p$ and $n$.

%

\section{Conclusion and discussions} \label{secconclusion}
We have shown theoretically and via simulation studies that:
\begin{itemize}
\item the \textit{limiting spectral distribution} (LSD) of RCV matrix
depends not only on that of the ICV matrix, but also on the
time-variability of covolatility process;

\item in particular, even with the same targeting ICV matrix,
the \textit{empirical spectral distribution} (ESD) of RCV matrix can vary a lot,
depending on how the underlying covolatility process evolves over
time;
\item for a class $\mathcal{C}$ of processes, our proposed
estimator, the \textit{time-variation adjusted realized covariance} (TVARCV)
matrix, possesses the following desirable properties as an estimator of
the ICV matrix: as long as the targeting
ICV matrix is the same, the ESDs of TVARCV matrices
estimated from processes with different covolatility paths will be
close to each other, sharing a unique limit; moreover, the LSD of
TVARCV matrix is related to that of the
targeting ICV matrix through the same
Mar\u{c}enko--Pastur equation as in the sample covariance matrix case.
\end{itemize}

Furthermore, we establish a Mar\u{c}enko--Pastur type theorem for
weighted sample covariance matrices. For a class $\mathcal{C}$ of
processes, we also establish a~Mar\u{c}enko--Pastur type theorem for
RCV matrices, which explicitly demonstrates how the time-variability of
the covolatility process affects the LSD of RCV matrix.

In practice, for given $p$ and $n$, based on the (observable) ESD of
TVARCV matrix, one can use existing algorithms to obtain an estimate of the
ESD of ICV matrix, which can then be applied to further
applications such as portfolio allocation, risk management,
etc.\vspace*{-2pt}

\section*{Acknowledgments}

We are very grateful to the Editor, the~Associate~Editor and anonymous
referees for their very valuable comments and \mbox{suggestions}.

\vspace*{-2pt}
\begin{supplement}[id=suppA]
\stitle{Supplement to ``On the estimation of integrated covariance
matrices of high
dimensional diffusion processes''}
\slink[doi]{10.1214/11-AOS939SUPP} 
\sdatatype{.pdf}
\sfilename{aos939\_supp.pdf}
\sdescription{This material contains the proof of Proposition \ref
{propsubclassc},
a detailed explanation of the second statement in
Remark \ref{rmkthm2}, and the proofs of the various lemmas in Section
\ref{ssecprelim} and Proposition \ref{propRVLDP}.}
\end{supplement}


\printaddresses

\end{document}